\documentclass[10pt]{article}
\usepackage{a4wide}
\usepackage{latexsym}
\usepackage{amsfonts}
\usepackage{amssymb}
\def\l{\langle}
\def\r{\rangle} 

\def\bq{\begin{eqnarray}}
\def\eq{\end{eqnarray}}
\def\bs{\begin{small}}
\def\es{\end{small}}
\newcounter{exercise}
\begin{document}

\thispagestyle{empty}

\begin{flushright}
NIKHEF-00-012
\end{flushright}

\vspace{1.5cm}

\begin{center}
  {\Large \bf Introduction to Monte Carlo methods}\\[.3cm]
  \vspace{1.7cm}
  {\sc Stefan Weinzierl$^{1}$}\\
  \vspace{1cm}
  {\it NIKHEF Theory Group\\
    Kruislaan 409, 1098 SJ Amsterdam, The Netherlands} \\
\end{center}

\vspace{2cm}

% abstract ---------------------------------------
\begin{abstract}\noindent
  {These lectures given to graduate students in high energy physics, 
   provide an introduction to Monte Carlo methods.
   After an overview of classical numerical quadrature rules, Monte Carlo integration
   together with variance-reducing techniques is introduced. 
   A short description on the generation of pseudo-random numbers and quasi-random numbers
   is given.
   Finally, methods to generate samples according to a specified distribution are discussed.
   Among others, we outline the Metropolis algorithm 
   and give an overview of existing algorithms for the generation
   of the phase space of final state particles in high energy collisions.
  }
\end{abstract}

\vspace*{\fill}

% footnotes -------------------------------------
 \noindent 
 $^1${\small email address : stefanw@nikhef.nl}

% main text ------------------------------------
\newpage
\tableofcontents
\newpage

\reversemarginpar

\section{Introduction}

Monte Carlo methods find application in a wide field of areas, including many subfields
of physics, like statistical physics or high energy physics, and ranging to areas
like biology or analysis of financial markets.
Very often the basic problem is to estimate a multi-dimensional integral
\bq
I & = & \int dx \; f(x)
\eq
for which an analytic answer is not known.
To be more precise one looks for an algorithm which gives a numerical estimate of 
the integral together with an estimate of the error. Furthermore the algorithm
should yield the result in a reasonable amount of time, e.g. at low computational
cost.
There is no point in developing an algorithm which gives the correct result but takes
forever.
However, there is no ``perfect'' algorithm suitable for all problems.
The most efficient algorithm depends on the specific problem. The more one knows
about a specific problem, the higher the chances to find an algorithm which solves
the problem efficiently.
We discuss in these lectures therefore a variety of methods, together with their
advantages and disadvantages and hope that the reader will gain experience where
a specific method can be applied.
We will foccus on the evaluation of multi-dimensional integrals as a guideline
through these lectures.  
Other applications of the Monte Carlo method, for example to optimization problems
are not covered in these lectures.
In the first section we discuss classical numerical quadrature rules. Monte Carlo
integration, together with various variance-reduction techniques is introduced
in the second section.
The third section is devoted to the generation of uniform pseudo-random numbers
and to the generation of quasi-random numbers in a $d$-dimensional hypercube.
The fourth section deals with the generation of samples according to a specified
probability distribution. This section is more focused on applications than the previous
ones. In high energy physics Monte Carlo methods are mainly applied in lattice calculations,
event generators and perturbative $\mbox{N}^n\mbox{LO}$-programs.
We therefore discuss the Metropolis algorithm relevant for lattice calculations, as well
as various methods how to generate four-vectors of final state particles in high energy
collisions according to the phase-space measure, relevant to event generators and
$\mbox{N}^n\mbox{LO}$-programs.

\section{Classical numerical integration}

Numerical quadrature rules have been known for centuries. They fall broadly in
two categories: Formulae which evaluate the integrand at equally spaced
abscissas (Newton-Cotes type formulae) and formulae which evaluate
the integrand at carefully selected, but non-equally spaced abscissa 
(Gaussian quadrature rules). The latter usually give better results
for a specific class of integrands.
We also discuss the Romberg integration technique as an example
of an extrapolation method. In the end of this section
we look at the deficiances which occur when numerical quadrature rules are
applied to multi-dimensional integrals.

\subsection{Newton-Cotes type formulae}

Formulae which approximate an integral over a finite interval by weighted values
of the integrand at equally spaced abscissas are called formulae of Newton-Cotes type.
The simplest example is the trapezoidal rule:
\bq
\int\limits_{x_0}^{x_0+\Delta x} dx f(x) & = & \frac{\Delta x}{2} 
\left[ f(x_0) + f(x_0+\Delta x) \right]
- \frac{(\Delta x)^3}{12} f''(\xi),
\eq
where $x_0 \le \xi \le x_0+\Delta x$.
To approximate an integral over a finite interval $[x_0,x_n]$ with the help of this formula
one divides the intervall into $n$ sub-intervals of length $\Delta x$ and applies the
trapezoidal rule to each sub-intervall. With the notation $x_j=x_0+j\cdot \Delta x$
one arrives at the compound formula
\bq
\label{trapez}
\int\limits_{x_0}^{x_n} dx f(x) & = & \frac{x_n-x_0}{n} \sum\limits_{j=0}^n w_j f(x_j)
- \frac{1}{12} \frac{(x_n-x_0)^3}{n^2} \tilde{f}''
\eq
with $w_0=w_n=1/2$ and $w_j=1$ for $1 \le j \le n-1$. Further
\bq
\tilde{f}'' & = & \frac{1}{n} \sum\limits_{j=1}^n f''(\xi_j),
\eq
where $\xi_j$ is somewhere in the intervall $[x_{j-1},x_j]$. 
Since the position of the $\xi_j$ cannot be known without knowing the integral
exactly, the last term in eq.~\ref{trapez} is usually neglected
and introduces an error in the numerical evaluation. Note that the error is proportional to $1/n^2$
and that we have to evaluate the function $f(x)$ roughly $n$-times (to be exact $(n+1)$-times,
but for large $n$ the difference does not matter).\\
\\
An improvement is given by Simpson's rule, which evaluates the function at three points:
\bq
\int\limits_{x_0}^{x_2} dx f(x) & = & 
\frac{\Delta x}{3}\left[ f(x_0) + 4 f(x_1) + f(x_2) \right]
-\frac{(\Delta x)^5}{90} f^{(4)}(\xi).
\eq
This yields the compound formula
\bq
\int\limits_{x_0}^{x_n} dx f(x) & = & 
\frac{x_n-x_0}{n} \sum\limits_{j=0}^n w_j f(x_j)
- \frac{1}{180} \frac{(x_n-x_0)^5}{n^4} \tilde{f}^{(4)},
\eq
where $n$ is an even number, $w_0=w_n=1/3$, and for $1\le j \le n$ we have
$w_j = 4/3$ if $j$ is odd and $w_j = 2/3$ if $j$ is even.  
The error estimate scales now as $1/n^4$.\\
\\
Newton's $3/8$-rule, which is based on the evaluation of the integrand 
at four points, does not lead to an improvement in the error estimate:
\bq
\int\limits_{x_0}^{x_3} dx f(x) & = & 
\frac{3 \Delta x}{8} \left[ f(x_0)+3f(x_1)+3f(x_2)+f(x_3) \right]
- \frac{3 (\Delta x)^5}{80} f^{(4)}(\xi).
\eq
As Simpson's rule it leads to a scaling of the error proportional to
$1/n^4$. An improvement is only obtained by going to a formula
based on five points (which is called Boole's rule):
\bq
\int\limits_{x_0}^{x_4} dx f(x) & = & 
\frac{2 \Delta x}{45} \left[
 7 f(x_0) + 32 f(x_1) + 12 f(x_2) + 32 f(x_3) + 7 f(x_4) \right]
-\frac{8 (\Delta x)^7}{945} f^{(6)}(\xi). \nonumber \\
\eq
Here the error of a compound rule scales like $1/n^6$.
In general one finds for an integration rule of
the Newton-Cotes type that, if the number of points in the starting formula is odd, say $2k-1$,  
then the error term is of the form
\bq
c_{2k-1} (\Delta x)^{2k+1} f^{(2k)}(\xi),
\eq
where $c_{2k-1}$ is some constant.
If the number of points is even, say $2k$, the error is of the same form:
\bq
c_{2k} (\Delta x)^{2k+1} f^{(2k)}(\xi).
\eq
A formula whose remainder term is proportional to $f^{(n+1)}(\xi)$
is called of degree $n$. Such a formula is exact for polynomials
up to degree $n$.
Now immediately one question arrises: What hinders us to improve
the error estimate for a compound rule based on Newton-Cotes formulae
by using even higher point formulae?
First, the more refined a rule is, the more
certain we must be that it is applied to a function which
is sufficiently smooth.
In the formulae above it is implied that, if the error term is proportional to $f^{(2k)}(\xi)$
that the function $f(x)$ is at least $(2k)$-times differentiable
and that $f^{(2k)}(x)$ is continous.
Applying the rule to a function which does not satisfy the criteria
can lead to a completely wrong error estimate.
(Try to apply the trapezoidal rule on the intervall $[0,1]$ to the function
$f(x)=0$ for $x<1/4$, $f(x)=1$ for $1/4<x<3/4$ and $f(x)=0$ for $x>3/4$.)
Secondly, it can be shown if the number of points become large,
the coefficients of the Newton-Cotes formulae become large and of
mixed sign. This may lead to significant numerical cancellations between
different terms and Newton-Cotes rules of increasing order
are therefore not used in practice.

\subsection{Gaussian quadratures}

The Newton-Cotes type rules approximate an integral of a function by the sum
of its functional values at a set of equally spaced points, multiplied
by appropriately chosen weights. We saw that as we allowed ourselves the freedom
in choosing the weights, we could achieve integration formulas of higher and higher
order.
Gaussian integration formulae take this idea further and allows us not only the freedom
to choose the weights appropriately, but also the location of the abscissas at which the function
is to be evaluated.
A Newton-Cotes type formula, which is based on the evaluation at $n$ points
is exact for polynomials up to degree $n$ (if $n$ is odd) or degree $(n-1)$
(if $n$ is even).
Gaussian quadrature formulae yield integration rules of degree $(2n-1)$.
Furthermore these rules can be generalized such that they don't yield
exact results for a polynomial up to degree $(2n-1)$, but for an integrand
of the form ``special function'' times ``polynomial up to degree $(2n-1)$''.\\
\\
Before stating the main formula for Gaussian quadratures we first
give an excursion to Lagrange's interpolation formula and introduce orthogonal
polynomials.

\subsubsection{Lagrange interpolation formula}

Let $x_0$, $x_1$, ..., $x_n$ be $(n+1)$ pairwise distinct points and let there be given
$(n+1)$ arbitrary numbers $y_0$, $y_1$, ..., $y_n$.
We wish to find a polynomial $p_n(x)$ of degree $n$ such that
\bq
p_n(x_i) = y_i, & & i=0,1,...,n.
\eq
The solution is due to Lagrange and is given by
\bq
p_n(x) & = & \sum\limits_{i=0}^n y_i l^n_i(x),
\eq
where the fundamental Lagrange polynomials are given by
\bq
l^n_i(x) & = & \frac{(x-x_0)...(x-x_{i-1})(x-x_{i+1})...(x-x_n)}
{(x_i-x_0)...(x_i-x_{i-1})(x_i-x_{i+1})...(x_i-x_n)}.
\eq
\bs
{\it Exercise \theexercise: Prove this statement. \\
Hint: You may first show that $l^n_i(x_j)$ equals 1 
if $i=j$ and zero if $i \neq j$.
\stepcounter{exercise}}
\es
\\
\\
If we want to approximate a function $f(x)$ by Lagrange's interpolating polynomial
$p_n(x)$ such that $f(x_i)=p(x_i)$ for $i=0,1,...,n$ the remainder term is given by
\bq
f(x) & = & p_n(x) + \frac{f^{(n+1)}(\xi)}{(n+1)!} \Pi(x),
\eq
where $\Pi(x)=(x-x_0)(x-x_1)...(x-x_n)$ and
$\mbox{min}(x,x_0)<\xi<\mbox{max}(x,x_n)$.
\\
\\
\bs
{\it Exercise \theexercise: Show that
\bq
p_{(2n+1)}(x) & = & 
\sum\limits_{i=0}^n \left[ f(x_i) \left( 1 -\frac{\Pi''(x_i)}{\Pi'(x_i)} (x-x_i) \right) 
+ f'(x_i) (x-x_i) \right] \left( l_i^n(x) \right)^2
\eq
is the unique polynomial of degree $(2n+1)$ for which
\bq
& & p_{(2n+1)}(x_i) = f(x_i), \;\;\; p_{(2n+1)}'(x_i) = f'(x_i) \;\;\;\mbox{for}\;\; i=0,1,...,n.
\eq
The remainder term is given by
\bq
f(x) & = & p_{(2n+1)}(x) + \frac{f^{(2n+2)}(\xi)}{(2n+2)!} \left( \Pi(x) \right)^2,
\eq
where $\mbox{min}(x,x_0)<\xi<\mbox{max}(x,x_n)$.
\stepcounter{exercise}}
\es

\subsubsection{Orthogonal polynomials}

A sequence of polynomials $P_0(x)$, $P_1(x)$, ..., in which $P_n(x)$ is of degree $n$
is called orthogonal with respect to the weight function $w(x)$ if
\bq
\int\limits_a^b dx \; w(x) P_i(x) P_j(x) = 0 & & \mbox{for} \; i \neq j.
\eq
Here we should mention that a function $w(x)$ defined on an interval $[a,b]$ is called
a weight function if $w(x) \ge 0$ for all $x\in [a,b]$, $\int\limits_a^b dx \; w(x) > 0$
and $\int\limits_a^b dx \; w(x) x^j < \infty$ for all $j=0,1,2,...$.
By rescaling each $P_n(x)$ with an appropriate constant one can produce a set of
polynomials which are orthonormal.
An important theorem states that the zeros of (real) orthogonal polynomials are real, simple
and located in the interior of $[a,b]$.
A second theorem states that if $x_1<x_2<...<x_n$ are the zeros of the orthogonal polynomial
$P_n(x)$, then in each interval $[a,x_1]$, $[x_1,x_2]$, ..., $[x_{n-1},x_n]$, $[x_n,b]$ there
is precisely one zero of the orthogonal polynomial $P_{n+1}(x)$.
Well-known sets of orthogonal polynomials are the Legendre,
Tschebyscheff, Gegenbauer, Jacobi, Laguerre and Hermite polynomials.
Some of them are generalizations or specializations of others.
In order to distinguish them one characterizes them by the interval on which they are
defined and by the corresponding weight function.
The first four (Legendre, Tschebyscheff, Gegenbauer and Jacobi) are defined on the 
intervall $[-1,1]$, the Laguerre polynomials are defined on $[0,\infty]$ and finally
the Hermite polynomials are defined on $[-\infty,\infty]$.
The weight function for the Legendre polynomials is simply $w(x)=1$, the weight
function for the Gegenbauer polynomials is $w(x) = (1-x^2)^{\mu-1/2}$ where
$\mu>-1/2$. The special cases corresponding to $\mu=0$ and $\mu=1$ give the
Tschebyscheff polynomials of the first and second kind, respectively.
And of course, in the case $\mu=1/2$ the Gegenbauer polynomials reduce
to the Legendre polynomials.
The Jacobi polynomials have the weight function $w(x)=(1-x)^\alpha (1+x)^\beta$ with
$\alpha,\beta>-1$. Here the special case $\alpha=\beta$ yields up to normalization
constants the Gegenbauer polynomials.
The generalized Laguerre polynomials have the weight function $w(x)=x^\alpha e^{-x}$
where $\alpha>-1$. Here $\alpha=0$ corresponds to the original Laguerre polynomials.
The Hermite polynomials correspond to the weight function $\exp(-x^2)$.
We have collected some formulae relevant to orthogonal polynomials in appendix \ref{orthogonal}.

\subsubsection{The main formula of Gaussian quadrature}

If $w(x)$ is a weight function on $[a,b]$, then there exists weights $w_j$ and abscissas $x_j$
for $1\le j \le n$ such that
\bq
\int\limits_a^b dx \;w(x) f(x) & = & 
\sum\limits_{j=1}^n w_j f(x_j) + \frac{f^{(2n)}(\xi)}{(2n)!}
\int\limits_a^b dx \;w(x) \left[ \Pi(x) \right]^2
\eq
with
\bq
& & \Pi(x) = (x-x_1) (x-x_2) ... (x-x_n), \nonumber \\
& & a \le x_1 < x_2 < ... < x_n \le b, \;\;\;\; a < \xi < b.
\eq
The abscissas are given by the zeros of the orthogonal polynomial of degree
$n$ associated to the weight function $w(x)$.
In order to find them numerically it is useful to know that they all lie in the interval $[a,b]$.
The weights are given by the (weighted) integral over the
Lagrange polynomials:
\bq
\label{gaussianweights}
w_j & = & \int\limits_a^b dx \; w(x) l^n_j(x).
\eq
\bs
{\it Exercise \theexercise: 
Let $P_0(x)$, $P_1(x)$, ... be a set of orthonormal polynomials, e.g.
\bq
\int\limits_a^b dx \; w(x) P_i(x) P_j(x) & = & \delta_{ij},
\eq
let $x_1$, $x_2$, ..., $x_{n+1}$ be the zeros of $P_{n+1}(x)$ and $w_1$, $w_2$, ..., $w_{n+1}$
the corresponding Gaussian weights given by eq.~\ref{gaussianweights}.
Show that for $i,j<n+1$
\bq
\sum\limits_{k=1}^{n+1} w_k P_i(x_k) P_j(x_k) & = & \delta_{ij},
\eq
e.g. the $P_0$, $P_1$, ..., $P_n$ are orthonormal on the zeros of $P_{n+1}$.
This equation can be useful to check the accuracy with which the zeros and weights of $P_{n+1}$ have been determined numerically.
\stepcounter{exercise}}
\es

\subsection{Romberg integration}

Suppose we have an integration rule of degree $(r-1)$ which evaluates the
integrand at $n$ points, e.g.
\bq
\label{originalrule}
\int\limits_a^b dx \; f(x) = S_{[a,b]}[f] + R_{[a,b]}[f], & &
S_{[a,b]}[f] = \sum\limits_{j=1}^n w_j f(x_j)
\eq
and $R_{[a,b]}[f]$ denotes the remainder. We may then construct
a new rule of degree $r$ as follows:
\bq
\label{newrule}
\int\limits_a^b dx \; f(x) & = & p S_{[a,b]}[f]
+ q \left( S_{[a,(a+b)/2]}[f] + S_{[(a+b)/2,b]} \right) + \tilde{R}_{[a,b]}[f]
\eq
with $p=-1/(2^r-1)$ and $q=2^r/(2^r-1)$. 
Since the original rule eq.~\ref{originalrule} is of degree $(r-1)$, the
remainder term $R_{[a,b]}[f]$ is of the form $c' (b-a)^{r+1} f^{(r)}(\xi)$,
but for $f=x^r$ the $r$-th derivative is a constant. We find therefore
for 
\bq
\tilde{R}_{[a,b]}[x^r] & = & p c (b-a)^{r+1} 
+ 2 q c \left(\frac{b-a}{2}\right)^{r+1} = 0.
\eq
This proves that the new rule eq.~\ref{newrule} is now of degree $r$.\\
\\
\bs
{\it Exercise \theexercise:
Take as the original rule the trapezoidal rule and construct the improved
rule. What do you get?
Repeat the exercise with Simpson's rule and Newton's 3/8 rule as starting
point. At how many different points do you have to evaluate the integrand
with the improved rules? Is this efficient? 
Can the fact that $p$ and $q$ have opposite signs cause any problems?
\stepcounter{exercise}}
\es
\\
\\
In practice Romberg integration is used with the trapezoidal rule.
Let
\bq
S_0^{(k)} & = & \frac{b-a}{2^k} \sum\limits_{j=0}^{2^k} w_j f(x_j)
\eq
be the trapezoidal sum with $2^k+1$ points and define
\bq
S_m^{(k)} & = & \frac{4^mS_{m-1}^{(k+1)}-S_{m-1}^{(k)}}{4^m-1}.
\eq
Then the series $S_0^{(0)}$, $S_1^{(0)}$, $S_2^{(0)}$ ...
converges better then the series
$S_0^{(0)}$, $S_0^{(1)}$, $S_0^{(2)}$ ....
Romberg integration is a special case of an extrapolation method 
(Richardson extrapolation to be precise). Based on a few estimates 
$S_0^{(k-i)}$, ..., $S_0^{(k)}$ with $2^{k-i}+1$, ..., $2^{k}+1$
points one extrapolates to the limit $k \rightarrow \infty$.

\subsection{Multi-dimensional integration}

In many problems multi-dimensional integrals occur, which have to be 
evaluated numerically.
One may try to extend the one-dimensional integration formulae to
$d$-dimensional integration formulae by viewing the $d$-dimensional
integral as an iteration of one-dimensional integrals and applying
a one-dimensional integration rule in each iteration.
As an example we consider an integral over the $d$-dimensional hypercube
$[0,1]^d$ evaluated with the help of the trapezoidal rule:
\bq
\int d^du \; f(u_1,...,u_d) & = & \frac{1}{n^d} \sum\limits_{j_1=0}^n ...
\sum\limits_{j_d=0}^n w_{j_1} ... w_{j_d} 
f\left(\frac{j_1}{n},...,\frac{j_d}{n}\right) + O\left(\frac{1}{n^2}\right).
\eq
In total we have to evaluate the function $N = (n+1)^d \approx n^d$ times.
Since the necessary computing time is proportional to $N$ we observe
that the error scales as $N^{-2/d}$. With increasing dimension $d$
the usefulness of the error bound $O(N^{-2/d})$ declines drastically.
Changing for example from the trapezoidal rule to Simpson's rule
does not change the situation significantly: The error bound
would scale in this case as $N^{-4/d}$.
We will later see that Monte Carlo integration yields an error which
decreases with $1/\sqrt{N}$ independent of the number of dimensions.
It has therefore become the method of choice for numerical integration
in high dimensions.

\subsection{Summary on numerical quadrature rules}

Numerical quadrature rules are the best method for one-dimensional
integrals. If the integrand is sufficiently smooth and if one knows an
absolute bound for a 
certain derivative of the integrand, they yield an exact error
estimate. 
The efficiency of numerical quadrature rules decreases rapidly with the
number of dimensions.
Furthermore, for complicated integration boundaries, which have to be
imbedded for example into a hypercube, the integrand is no longer
a smooth function and the estimate for the error can no longer
be used.\\
\\
Further reading: You can find more information on numerical quadrature rules
in the books by Davis and Rabinowitz \cite{Davis} and by Press et al. \cite{Recipes}.
\\
\\
\bs
{\it Exercise \theexercise: Evolution of parton densities using
numerical quadrature with Laguerre polynomials.
This method is due to D.A. Kosower \cite{kosower}.
The evolution equations for the quark non-singlet parton distributions $f(x,Q^2)$ of the proton
read
\begin{eqnarray}
\label{evolution}
Q^2 \frac{\partial f(x,Q^2)}{\partial Q^2} & = & P(x,Q^2) \otimes f(x,Q^2),
\end{eqnarray}
where $x$ stands for the nucleon's momentum fraction carried by the parton,
$P(x,Q^2)$ is the Altarelli-Parisi evolution kernel, and $\otimes$
denotes the convolution
\begin{eqnarray}
A(x) \otimes B(x) & = & \int\limits_0^1 dy \int\limits_0^1 dz.
\delta(x- y z) A(y) B(z) 
\end{eqnarray}
The evolution kernel is known to next-to-leading order:
\begin{eqnarray}
P(x,Q^2) & = & a_s(Q^2) P_0(x) + a_s^2(Q^2) P_1(x) + O(a_s^3),
\end{eqnarray}
where we have introduced $a_s(Q^2) = \alpha_s(Q^2)/4\pi$.
In the Mellin-transform approach one factorizes eq.~\ref{evolution} by taking Mellin 
moments.
The Mellin moments of a function $h(x)$ are given by
\bq
h^z & = & \int\limits_0^1 dx \; x^{z-1} h(x).
\eq
Truncating to next-to-leading order we obtain:
\begin{eqnarray}
Q^2 \frac{\partial f^z(Q^2)}{\partial Q^2} & = & 
\left( a_s(Q^2) P_0^z + a_s^2(Q^2) P_1^z \right) f^z(Q^2). 
\end{eqnarray}
This equation is formally solved by the introduction of an evolution
operator $E^z$:
\begin{eqnarray}
f^z(Q^2) & = & E^z(Q^2,Q_0^2) f^z(Q_0^2).
\end{eqnarray}
The evolution operator is for the
quark non-singlet case of the form:
\begin{eqnarray}
E^z(Q^2,Q_0^2) & = & \left( \frac{a_s(Q^2)}{a_s(Q_0^2)}\right)^{\frac{\gamma^z_0}
{2 \beta_0}}
\left[ 1 + \frac{a_s(Q^2)-a_s(Q_0^2)}{2 \beta_0}
\left( \gamma^z_1 - \frac{\beta_1}{\beta_0} \gamma^z_0 \right) \right],
\end{eqnarray}
where $\gamma_0^z$ and $\gamma_1^z$ are the first and second coefficients of the anomalous
dimensions.
Retransforming back one obtains the evolved parton distribution in x-space by
\begin{eqnarray}
f(x,Q^2) & = & \frac{1}{2\pi i} \int\limits_C dz x^{-z} f^z(Q^2),
\end{eqnarray}
where the contour $C$ runs to the right of all singularities of the integrand.\\
The parton distributions are usually parametrized at the input scale $Q_0^2$ in a
form like
\begin{eqnarray}
x f(x,Q_0^2) & = & \sum\limits_{i} A_i x^{\alpha_i} (1-x)^{\beta_i} 
\end{eqnarray}
with Mellin transform
\begin{eqnarray}
f^z(Q_0^2) & = & \sum\limits_{i} A_i B(z+\alpha_i-1,1+\beta_i),
\end{eqnarray}
where $B(x,y)$ is the beta function.
One example is the CTEQ 4M structure function for the u-quark valence distribution $u_v$, given at the scale
$Q_0=1.6 \mbox{GeV}$:
\begin{eqnarray}
\label{cteq}
x u_v & = & 1.344\; x^{0.501} (1-x)^{3.689} \left( 1 + 6.402 \; x^{0.873} \right).
\end{eqnarray}
Droping from now on the arguments $Q^2$ and $Q_0^2$ our task is to
evaluate the integral
\begin{eqnarray}
f(x,Q^2) = \frac{1}{\pi} \mbox{Re} \int\limits_{C_s}  dz \frac{1}{i} F(z), & &
F(z) = x^{-z} E^z \sum\limits_i A_i B(z+\alpha_i-1,1+\beta_i),
\end{eqnarray}
where we have used complex conjugation to replace the contour $C$ by $C_s$,
starting at the real axis, right to the right-most pole and running upwards
to infinity.
The most elegant way to solve this problem is to choose a contour in such a
way that the integrand can very well be approximated by some set of
orthogonal polynomials.
We neglect for the moment the evolution operator $E^z$.
We try a parabolic contour
\begin{eqnarray*}
z(t)  & =  & z_0 + i t + \frac{1}{2} c_3 t^2 
\end{eqnarray*}
and determine the parameters $z_0$ and $c_3$ according to
\bq
\label{z0}
F'(z_0)  =  0, & & c_3 = \frac{F'''(z_0)}{3 F''(z_0)}.
\eq
Finally we change variables
\bq
u = \left( \frac{t}{c_2} \right)^2, & & c_2 = \sqrt{\frac{2 F(z_0)}{F''(z_0)}}.
\eq
As the result we obtain
\begin{eqnarray}
f(x,Q^2) & = & \frac{c_2}{2 \pi} \int\limits_{0}^{\infty}
  \frac{du}{\sqrt{u}} e^{-u} \mbox{Re}
  \left[ e^{u} \left(1-i c_2 c_3 \sqrt{u} \right)
       F(z(c_2 \sqrt{u})) \right]
\end{eqnarray}
and the integrand can be approximated
by Laguerre polynomials $L_n^{-1/2}(x)$.
Start from the CTEQ parameterization and find $z_0$ by solving 
$F'(z_0)=0$ numerically 
(you may use Newton-Raphson for example) and determine the parameters $c_2$ and $c_3$.
Evaluate the integral by using a Gaussian quadrature formula for Laguerre polynomials with 
$3$, $5$ or $10$ points.
Here you have to find the correct abscissas and weights.
Since we set the evolution operator $E^z=1$ you should recover the original parton density
at the scale $Q_0^2$. 
The inclusion of the evolution operator does only slighty modify the integrand and one
can therefore use the same contour as in the non-evolved case.
(The relevant anomalous dimensions can be found for example in \cite{floratos}.)
\stepcounter{exercise}}
\es

\section{Monte Carlo techniques}

We have seen in the previous section that numerical quadrature rules are
inefficient for multi-dimensional integrals. In this section we introduce
Monte Carlo integration. We will show that for Monte Carlo integration
the error scales like $1/\sqrt{N}$, independent of the number of dimensions.
This makes Monte Carlo integration the preferred method for integrals
in high dimensions. But, after the first euphoria is gone, one
realizes that convergence by a rate of $1/\sqrt{N}$ is pretty slow.  
We discuss therefore several techniques to improve the efficiency of
Monte Carlo integration.

\subsection{Monte Carlo integration}

We consider the integral of a function $f(u_1,...,u_d)$, depending on $d$ variables
$u_1$, ..., $u_d$ over the unit hypercube $[0,1]^d$. 
We assume that $f$ is square-integrable.
As a short-hand notation we will
denote a point in the unit hypercube by $x=(u_1,...,u_d)$ and the function evaluated
at this point by $f(x)=f(u_1,...,u_d)$.
The Monte Carlo estimate for the integral
\bq
\label{basicMCintegral}
I & = & \int dx f(x) = \int d^du f(u_1,...,u_d)
\eq
is given by
\bq
E & = & \frac{1}{N} \sum\limits_{n=1}^N f(x_n).
\eq
The law of large numbers ensures that the Monte Carlo estimate
converges to the true value of the integral:
\bq
\lim\limits_{N\rightarrow \infty} \frac{1}{N} \sum\limits_{n=1}^N f(x_n)
& = & I.
\eq
In order to discuss the error estimate for finite $N$, we first introduce
the variance $\sigma^2(f)$ of the function $f(x)$:
\bq
\sigma^2(f) & = & \int dx \left( f(x) - I \right)^2.
\eq
We can then show that
\bq
\label{variance}
\int dx_1 ... \int dx_N \left(
\frac{1}{N} \sum\limits_{n=1}^N f(x_n) - I \right)^2
& = & \frac{\sigma^2(f)}{N}.
\eq
\bs
{\it Exercise \theexercise:
Prove eq.~\ref{variance}. \\
Hint: You may want to introduce an auxiliary
function $g(x)=f(x)-I$ and show $\int dx g(x) = 0$ first.
\stepcounter{exercise}}
\es \\
\\
Eq.~\ref{variance} can be interpreted to mean that the error
in the Monte Carlo estimate is on the average $\sigma(f)/\sqrt{N}$.
$\sigma(f)$ is called the standard deviation of $f$.
The central limit theorem tells us then that the probability that our Monte
Carlo estimate lies between $I-a \sigma(f)/\sqrt{N}$ and 
$I+b \sigma(f)/\sqrt{N}$ is given by
\bq
\lim\limits_{N\rightarrow \infty} \; \mbox{Prob} \left(
-a \frac{\sigma(f)}{\sqrt{N}} \le 
\frac{1}{N} \sum\limits_{n=1}^N f(x_n) - I \le
b \frac{\sigma(f)}{\sqrt{N}} \right) & = & 
\frac{1}{\sqrt{2\pi}} \int\limits_{-a}^b dt \exp\left( -\frac{t^2}{2} \right).
\eq
This also shows that error in a Monte Carlo integration scales
like $1/\sqrt{N}$ independent of the dimension $d$.
Of course, in practice one cannot obtain easily the exact value for the
variance $\sigma^2(f)$ and one uses the Monte Carlo estimate
\bq
S^2 & = & \frac{1}{N-1} \sum\limits_{n=1}^N \left( f(x_n) - E \right)^2
= \frac{1}{N} \sum\limits_{n=1}^N \left( f(x_n) \right)^2 - E^2
\eq
instead.\\
\\
\bs
{\it Exercise \theexercise:
For a reliable error estimate we had to require that the function $f$
is square integrable. If the function $f$ is integrable, but not square
integrable, the Monte Carlo estimate $E$ for the integral will still
converge to the true value, but the error estimate will become unreliable.
Do a Monte Carlo integration of the function $f(x)=1/\sqrt{x}$ over
the interval $[0,1]$.
\stepcounter{exercise}}
\es
\\
\\
We would like to draw the attention to the fact that Monte Carlo
integration gives only a probabilistic error bound, e.g. we can only give
a probability that the Monte Carlo estimate lies within a certain range
of the true value. 
This should be compared to numerical quadrature formulae. If we use
for example the trapezoidal rule and if we know that the second derivative
of $f$ is bounded by, say
\bq
|f''(x)| \le c, & & x_0 \le x \le x_0+\Delta x,
\eq 
we obtain a deterministic error bound:
\bq
\left| 
\frac{\Delta x}{2}\left( f(x_0) + f(x_0+\Delta x) \right) 
- \int\limits_{x_0}^{x_0+\Delta x} dx \; f(x)
\right| \le \frac{c \left( \Delta x \right)^3}{12}.
\eq
Here the error is guaranteed to be smaller than $c (\Delta x)^3/12$. On the other hand
to obtain this bound we used the additional information on the second derivative.\\
\\
\bs
{\it Exercise \theexercise:
Buffon's needle technique to estimate $\pi$.
Buffon used in 1777 the following procedure to estimate $\pi$ : A pattern of parallel lines
separated by a distance $d$ is laid out on the floor. Repeatedly a needle of length $d$ is thrown
onto this stripped pattern. Each time the needle lands in such a way as to cross a boundary between
two stripes, it is counted as a hit. The value of $\pi$ is then estimated from twice the number
of tries divided by the number of hits.
The above recipe is based on the fact that the probability of a hit is $2/\pi$. This can be seen as
follows: Let $\varphi$ be the
angle between the needle and the perpendicular to the stripes. For a given $\varphi$ the probability
of a hit is $|\cos \varphi|$. Since all angles are equally likely, the avarage value of
$|\cos \varphi|$ can be calculated by integrating over all angles and dividing by the range.
The problem with that algorithm is, that it gives large statistical errors.
Write a Monte Carlo program which estimates $\pi$ by using Buffon's needle technique.
How often do you have to throw the needle to get the first three, five or seven digits of $\pi$? 
\stepcounter{exercise}}
\es

\subsection{Variance reducing techniques}

We have seen that the error estimate of a Monte Carlo integration scales like
$1/\sqrt{N}$. The main advantage of Monte Carlo integration is the fact, that the
error estimate is independent of the dimension $d$ of the integral.
However we pay the price that the Monte Carlo estimate for the integral converges
relatively slow to the true value at the rate of $1\sqrt{N}$.
Several techniques exist to improve the situation.
  
\subsubsection{Stratified sampling}

This technique consists in dividing the full integration space into subspaces, performing a Monte Carlo
integration in each subspace, and adding up the partial results in the end. Mathematically, this is 
based on the fundamental property of the Riemann integral 
\begin{eqnarray}
\label{stratified}
\int\limits_0^1 dx f(x) & = & \int\limits_0^a dx f(x) + \int\limits_a^1 dx f(x), \;\;\; 0 < a < 1.
\end{eqnarray}
More generally we split the integration region $M=[0,1]^d$ into $k$ regions
$M_j$ where $j=1,...,k$. In each region we perform a Monte Carlo integration
with $N_j$ points. For the integral $I$ we obtain the estimate
\bq
E & = & \sum\limits_{j=1}^k \frac{\mbox{vol}(M_j)}{N_j} \sum\limits_{n=1}^{N_j} f(x_{jn})
\eq
and instead of the variance $\sigma^2(f)/N$ we have now the expression
\bq
\sum\limits_{j=1}^k \frac{\mbox{vol}(M_j)^2}{N_j} \left. \sigma^2(f)\right|_{M_j}
\eq
with
\bq
\left. \sigma^2(f)\right|_{M_j} & = & 
\frac{1}{\mbox{vol}(M_j)} \int\limits_{M_j} dx \; \left(
f(x) - \frac{1}{\mbox{vol}(M_j)} \int\limits_{M_j} dx \; f(x) \right)^2 \nonumber \\
& = & \left( \frac{1}{\mbox{vol}(M_j)} \int\limits_{M_j} dx \; f(x)^2 \right)
-
\left( \frac{1}{\mbox{vol}(M_j)} \int\limits_{M_j} dx \; f(x) \right)^2.
\eq
If the subspaces and the number of points in each subspace are chosen carefully, this can lead to a dramatic
reduction in the variance compared with crude Monte Carlo, but it should be noted that it 
can also lead to a larger variance if the choice is not appropriate.
If we take $a=1/2$ in eq.~\ref{stratified} and use $N_a$ points in the first region $[0,a]$ and
$N_b$ points in the second region $[a,1]$
we obtain for the error estimate
\bq
\frac{1}{4} \left( \frac{\left. \sigma^2(f) \right|_a}{N_a} + \frac{\left. \sigma^2(f) \right|_b}{N_b} \right),
\eq
which is minimized for fixed $N=N_a+N_b$ by choosing
\bq
\frac{N_a}{N} & = & \frac{\left. \sigma(f) \right|_a}{\left. \sigma(f) \right|_a + \left. \sigma(f) \right|_b}.
\eq
In general the total variance is minimized when the number of points in each subvolume is
proportional to $\left. \sigma(f) \right|_{M_j}$.

\subsubsection{Importance sampling}

Mathematically, importance sampling corresponds to a change of integration variables :
\bq
\int dx \; f(x) = \int \frac{f(x)}{p(x)} p(x) dx = \int \frac{f(x)}{p(x)} dP(x)
\eq
with 
\bq
p(x) & = & \frac{\partial^d}{\partial x_1 ... \partial x_d} P(x).
\eq
If we restrict $p(x)$ to be a positive-valued function $p(x) \ge 0$ and to be normalized
to unity
\bq
\int dx \; p(x) & = & 1
\eq
we may interpreted $p(x)$ as a probability density function.
If we have at our disposal a random number generator corresponding to the distribution $P(x)$
we may estimate the integral from a sample $x_1$, ..., $x_N$ of random numbers distributed
according to $P(x)$:
\bq
E & = & \frac{1}{N} \sum\limits_{n=1}^N \frac{f(x_n)}{p(x_n)}.
\eq
The statistical error of the Monte Carlo integration is given by $\sigma(f/p)/\sqrt{N}$, where
an estimator for the variance $\sigma^2(f/p)$ is given by
\bq
S^2\left( \frac{f}{p} \right) & = & \frac{1}{N} \sum\limits_{n=1}^N \left(
\frac{f(x_n)}{p(x_n)} \right)^2 - E^2.
\eq
We see that the relevant quantity is now $f(x)/p(x)$ and it will be advantageous to choose
$p(x)$ as close in shape to $f(x)$ as possible.
If $f(x)$ is positive-valued one might be tempted to choose $p(x)=cf(x)$. The constant is easily
found to be $1/I$ and the variance $\sigma^2(f/p)$ turns out to be zero. So we would have found
a perfect method which would return the correct value with only one sampling point.
Of course life in not so simple and the crucial point is that in order to sample
$f/p$ we must know $p$, and in order to know $p(x)=f(x)/I$ we must know $I$, and if we already
know $I$ we don't need a Monte Carlo integration to estimate it.
So in practice one chooses $p(x)$ such that it approximates $|f(x)|$ reasonably well 
in shape and such
that one can generate random numbers distributed according to $P(x)$.\\
\\
One disadvantage of importance sampling is the fact, that it is dangerous 
to choose functions $p(x)$, which
become zero, or which approach zero quickly. If $p$ goes to zero somewhere where $f$ is not zero,
$\sigma^2(f/p)$ may be infinite and the usual technique of estimating the 
variance from the sample points
may not detect this fact if the region where $p=0$ is small.

\subsubsection{Control variates}

As in importance sampling one seeks an integrable function $g$ which approximates the function $f$ to
be integrated, but this time the two functions are subtracted rather than divided. Mathematically,
this technique is based on the linearity of the integral :
\begin{eqnarray}
\int dx \; f(x) &= & \int dx \left( f(x) - g(x) \right) + \int dx \; g(x).
\end{eqnarray}
If the integral of $g$ is known, the only uncertainty comes from the integral of $(f-g)$, which
will have smaller variance than $f$ if $g$ has been chosen carefully.
The method of control variates is more stable than importance sampling, since zeros in $g$ cannot induce
singularities in $(f-g)$. Another advantage over importance sampling is that the integral of the 
approximating function $g$ need not be inverted analytically.

\subsubsection{Antithetic variates}

Usually Monte Carlo calculations use random points, which are independent of each other. The method of
antithetic variates deliberately makes use of correlated points,
taking advantage of the fact that such a correlation may be negative.
Mathematically this is based on the fact that
\begin{eqnarray}
\mbox{var}(f_1+f_2) & = & \mbox{var}(f_1) + \mbox{var}(f_2) + 2 \; \mbox{covar}(f_1,f_2).
\end{eqnarray}
If we can arrange to choose points such that $f_1$ and $f_2$ are negative correlated, a substantial
reduction in variance may be realized. The most trivial example for the application of the method of
antithetic variates would be a Monte Carlo integration of the function $f(x) =x $ in the intervall
$[0,1]$ by evaluating the integrand at the points $x_i$ and $1-x_i$.

\subsection{Adaptive Monte Carlo methods}

The variance-reducing techniques described above require some advance knowledge of the behaviour
of the function to be integrated.
In many cases this information is not available and one prefers adaptive techniques, e.g. an algorithm
which learns about the function
as it proceeds.
In the following we describe the VEGAS-algorithm \cite{ran4,ran5}, which is widely used in high-energy physics. 
VEGAS combines the basic ideas of importance sampling and stratified sampling into
an iterative algorithm, which automatically concentrates evaluations of the 
integrand in those regions where
the integrand is largest in magnitude. 
VEGAS starts by subdividing the 
integration space into a rectangular grid and
performs an integration in each subspace.
These results are then used to adjust the grid for the next iteration, according to where
the integral receives dominant contributions. 
In this way VEGAS uses importance sampling and tries to approximate the optimal
probability density function
\bq
p_{optimal}(x) & = & \frac{|f(x)|}{\int dx |f(x)|}
\eq
by a step function.
Due to storage requirements one has to use a separable probability density function
in $d$ dimensions:
\bq
p(u_1,...,u_d) & = & p_1(u_1) \cdot p_2(u_2) \cdot ... \cdot p_d(u_d).
\eq
Eventually after a few iterations the optimal grid is found.
In order to avoid rapid destabilizing changes in the grid, the adjustment of the grid
includes usually a damping term.
After this initial exploratory phase, the grid may be frozen and in a second evaluation phase the integral
may be evaluated with high precision according to the optimized grid. The separation into an exploratory
phase and an evaluation phase allows one to use less integrand evaluations in the first phase and to
ignore the numerical estimates from this phase (which will in general have a larger variance).
Each iteration yields an estimate $E_j$ together with an estimate for the variance $S_j^2$:
\bq
E_j = \frac{1}{N_j} \sum\limits_{n=1}^{N_j} \frac{f(x_n)}{p(x_n)},
& & 
S_j^2 = \frac{1}{N_j} \sum\limits_{n=1}^{N_j} \left( \frac{f(x_n)}{p(x_n)} \right)^2 - E_j^2.
\eq
Here $N_j$ denotes the number of integrand evaluations in iteration $j$.
The results of each iteration in the evaluation phase are combined into a cummulative estimate, 
weighted by the number of calls $N_j$
and their variances:
\bq
E & = & \left( \sum\limits_{j=1}^m \frac{N_j}{S_j^2} \right)^{-1}
\left( \sum\limits_{j=1}^m \frac{N_j E_j}{S_j^2} \right).
\eq
If the error estimates $S_j^2$ become unreliable (for example if the function is not square 
integrable),
it is more appropriate to weight the partial results by the number $N_j$ of integrand evaluations alone.
In addition VEGAS returns the $\chi^2$ per degree of freedom:
\bq
\chi^2/\mbox{dof} & = & \frac{1}{m-1} \sum\limits_{j=1}^m \frac{(E_j-E)^2}{S_j^2}.
\eq
This allows a check whether the various estimates are consistent.
One expects a $\chi^2/\mbox{dof}$ not much greater of one.\\
\\
In low dimensions one may use stratified sampling instead of importance sampling. If in each cell
at least two points are thrown, the contribution to the variance from each cell may be estimated
and the grid can be adjusted such that it minimizes the total variance. This method is however
restricted to low dimensions. If $b$ denotes the number of bins per axis, $d$ the number of 
dimensions and one requires two points per cell, one needs at least
\bq
N & = & 2 b^d
\eq
integrand evaluations. This number grows exponentially with $d$ and for large $d$ one has to resort to
importance sampling.
Note that for large $d$ there are inevitably cells, into which no point is thrown.\\
\\
\bs
{\it Exercise \theexercise: Write a computer program which implements the adaptive grid
technique of the VEGAS-algorithm in one dimension.
It should integrate a function $f(x)$ in the intervall $[0,1]$. Split the interval
$[0,1]$ into $b$ bins $[x_{i-1},x_i]$, where $0=x_0<x_1<...<x_b=1$.
The probability that a point is thrown into bin $i$ is $1/b$. Inside a bin the points
a chosen with a uniform probability.
After each iteration the boundaries are readjusted. Adjust the bins in such a way that
each bin would give a contribution of $1/b \int dx |f(x)|$,
based on the estimate you got in the last iteration.
\stepcounter{exercise}}
\es

\subsection{Multi-channel Monte Carlo}

If the integrand $f(x)$ has sharp peaks, crude Monte Carlo usually leads to poor results.
The situation can sometimes be improved by a remapping of variables, such that the integrand
becomes more flat.
However there might be situations where the integrand exhibits different peaks
in different regions. In such cases it is often impossible to find a variable
transformation, which remaps all peaks simultaneously.
Multi-channel Monte Carlo offers a solution, if the transformations for a single
peak structure are known.
Each such transformation is known as a channel.
Each channel is specified by a probability density function $p_i(x)$ and
a mapping $P_i^{-1}(y)$ from random numbers $y$ distributed according to $p_i(x)$ into the 
region of integration:
\bq
x & = & P_i^{-1}(y).
\eq
Each density is non-negative and normalized to unity: $\int dx p_i(x) = 1$
for $i=1,...,m$, where $m$ denotes the number of channels.
Let $\alpha_i \ge 0$, $i=1,...,m$ be non-negative numbers, such that
\bq
\sum\limits_{i=1}^n \alpha_i & = & 1.
\eq
A specific channel is then selected with probability $\alpha_i$.
In practice one fixes the total number of integrand evaluations and 
evaluates each channel roughly $N_i \approx \alpha_i N$ times.
The integral we want to calculate is
\bq
I & = & \int dx \; f(x) = \sum\limits_{i=1}^m \alpha_i \int \frac{f(x)}{p(x)} dP_i(x),
\eq 
where
$p(x) = \sum \alpha_i p_i(x)$. The Monte Carlo estimate for the integral is then
\bq
E & = & \frac{1}{N} \sum\limits_{i=1}^m \sum\limits_{n_i=1}^{N_i} \frac{f(x_{n_i})}{p(x_{n_i})}.
\eq
The expected error of the integration is given by
\bq
\sqrt{\frac{W(\alpha)-I^2}{N}},
\eq
where $W(\alpha)$ is given by
\bq
W(\alpha) & = & \sum\limits_{i=1}^m \alpha_i \int \left( \frac{f(x)}{p(x)} \right)^2 dP_i(x).
\eq
By adjusting the parameters $\alpha_i$ one may try to minimize $W(\alpha)$.
Since the integral $I$ does not depend on the parameters $\alpha_i$ one may change the $\alpha_i$
during the integration. The $\alpha_i$ do not affect the estimate for the integral, but only
the estimate for the error.\\
\\
The method suggested in \cite{multichannel} starts from an initial set 
$\alpha_i'$, performs a few hundred Monte Carlo evaluations to estimate
\bq
W_i(\alpha') & = & \int dx \; p_i(x) \left(\frac{f(x)}{p(x)}\right)^2
\eq
and rescales the parameters according to 
\bq
\alpha_i & = & \frac{ \alpha_i' \left(W_i(\alpha') \right)^\beta}
{\sum_i \alpha_i' \left(W_i(\alpha') \right)^\beta}.
\eq
The suggested values for the parameter $\beta$ range from $1/2$ to $1/4$ \cite{multichannel2}.

\subsection{Summary on Monte Carlo techniques}

Monte Carlo integration offers a tool for numerical evaluation of integrals
in high dimensions. Furthermore Monte Carlo integration works for smooth
integrands as well as for integrands with discontinuities. This allows
an easy application to problems with complicated integration boundaries.
However the error estimate scales always with $1/\sqrt{N}$. To improve
the situation we introduced the classical variance reducing techniques
(stratified sampling, importance sampling, control variates, antithetic
variates) as well as two advanced methods: The adaptive VEGAS-algorithm,
which learns about the integrand as it proceeds and multi-channel 
Monte Carlo, which is useful when the integrand is sharply peaked
in different regions of the integration region.\\
\\
Further reading: The material presented in this section is based on the book
by Hammersley and Handscomb \cite{hammersley} and the review article
by James \cite{ran1}.

\section{Random numbers}

Since a computer is a deterministic machine, truly random numbers do not exist
on a computer. One uses therefore pesudo-random numbers.
Pseudo-random numbers are produced in the computer deterministicly by a simple algorithm,
and are therefore not truly random, but any sequence of pseudo-random
numbers is supposed to appear random to someone who doesn't know
the algorithm. More quantitatively one performs for each
proposed pseudo-random number generator 
a series of tests $T_1$, $T_2$, ..., $T_n$.
If the outcome of one test differs significantly from what one would expect
from a truly random sequence, the pseudo-random number generator is classified
as ``bad''. Note that if a pseudo-random number generator has passed $n$ tests,
we can not conclude that it will also pass test $T_{n+1}$. 
\\
In this context also the term ``quasi-random numbers'' appears.
Quasi-random numbers are not random at all, but produced by a numerical algorithm and 
designed to be distributed
as uniformly as possible, in order to reduce the errors in Monte
Carlo integration.

\subsection{Pseudo-random numbers}

By today's standard a good random number generator should satisfy 
the following criteria:
\begin{itemize}
\item Good distribution. The points should be distributed
according to what one would expect from a truly random distribution.
Furthermore a pseudo-random number generator should not
introduce artificial correlations between succesivley generated points. 
\item Long period. Both pseudo-random and quasi-random generators always
have a period, after which they begin to generate the same sequence
of numbers over again. To avoid undesired correlations one should in any
practical calculation not come anywhere near exhausting the period.
\item Repeatability. For testing and development, it may be necessary
to repeat a calculation with exactly the same random numbers as in the
previous run. Furthermore the generator should allow the possibility
to repeat a part of a job without doing the whole thing. This requires
to be able to store the state of a generator.
\item Long disjoint subsequences. For large problems it is extremely
convenient to be able to perform independent subsimulations whose
results can later be combined assuming statistical indepedence.
\item Portability. This means not only that the code should be portable
(i.e. in a high-level language like Fortran or C), but that it should
generate exactly the same sequence of numbers on different machines.
\item Efficiency. The generation of the pseudo-random numbers should
not be too time-consuming. Almost all generators can be implemented 
in a reasonably efficient way. 
\end{itemize}
To test the quality of a pseudo-random number generator one performs
a series of test.
The simplest of all is the frequency test. If the algorithm claims to
generate pseudo-random numbers uniformly distributed in $[0,1]$, one divides
this intervall into $b$ bins, generates $n$ random number $u_1$, $u_2$, ...
$u_n$ and counts the number a pseudo-random number falls into bin $j$ ($1\le j \le b$).
One then calculates $\chi^2$ assuming that the numbers are truly random and obtains
a probability that the specific generated distribution is compatible
with a random distribution.
The serial test is a generalization of the frequency test. Here one looks at pairs of succesive
generated numbers and checks if they are uniformly distributed in the area $[0,1] \times [0,1]$.\\
\\
Another popular test is the gap test. Let $\alpha$ and $\beta$ be two real numbers
with $0 \le \alpha < \beta \le 1$. Given a sequence $u_1$, $u_2$, ... 
of supposedly random numbers
one determines the length of consecutive subsequences $u_j$, $u_{j+1}$, ...,
$u_{j+r}$ in which  $u_{j+r}$ lies between $\alpha$ and $\beta$, but the other
$u$'s don't.
Such a subsequence is called a gap of length $r$.
Having determined the gaps of length $0$, $1$, ... ,$t$, one then applies a $\chi^2$-test
to this empirical distribution.  \\
\\
We give an overview of some of the most popular algorithms for pseudo-random number generators.
All algorithm are given such that they generate integer numbers up to $m$. To obtain real
numbers in the interval $[0,1]$ one divides therefore by $m$.
Some algorithm can be implemented in such a way that they work directly with floating
point numbers.

\subsubsection{Multiplicative linear congruential generator}

Each succesive integer is obtained by multiplying the previous one
by a well chosen multiplier, optionally adding another constant, and
throwing away the most significant digits of the result:
\begin{eqnarray}
s_{i} & = & ( a s_{i-1} + c ) \;\mbox{mod}\; m.
\end{eqnarray}
where $a$ and $m$ are relativly prime. 
The linear congruential generator has a period no longer than $m$. 
Its theory is quite well understood \cite{ran3}.
For the modulus $m$ one chooses usually a number of the form $2^r$, $2^r+1$ or
$2^r-1$, which allow a fast implementation of the routine.
The choice $m=2^r$ is not always the best one. In this case it can be shown
that the $l$ lowest order bits of $s_i$ have a period not exceeding $2^l$.
For example, the last bit is either constant or stricly alternating, the last
two bits have a period no longer than two, the last three bits a period no longer
than 8.
In most applications the lowest order bits are however insignificant. 
As already stated the maximal period is $m$. There is a theorem which states
that the generator has the maximal period if and only if
$c>0$ is relatively prime to $m$, $a-1$ is a multiple of $p$ for every prime $p$
dividing $m$ and $a-1$ is a multiple of $4$, if $4$ divides $m$.
In the case $c=0$ the maximal possible period is attained if
$s_0$ is relatively prime to $m$ and if $a$ is a primitive element modulo $m$.
(If $a$ is relatively prime to $m$, the smallest integer $\lambda$ for which
$a^\lambda=1 \; \mbox{mod}\;m$ is called the order of $a$ modulo $m$. Any such $a$
which has the largest possible order modulo $m$ is called a primitive element modulo $m$.)
In this case the period is $\lambda(2^r)=2^{r-2}$ if $m=2^r$ with $r\ge3$, 
$\lambda(p^r)=p^{r-1} (p-1)$ if $m=p^r$ with $p>2$ a prime number
and the least common multiple of $\lambda(p_1^{r_1})$, ...
$\lambda(p_t^{r_t})$ if $m=p_1^{r_1} \cdot ... \cdot p_t^{r_t}$ with $p_1$, ..., $p_t$ being prime
numbers.
This leaves the question in which cases $a$ is a primitive element modulo $m$. 
If $m=2^r$ with $r\ge 4$ the answer is if and only if $a \; \mbox{mod} \; 8$ equals $3$ or $5$.
In the case $m=p$ where $p$ is a prime number greater than $2$, $a$ has to satisfy
$a \neq 0 \; \mbox{mod} \; p$ and $a^{(p-1)/q} \neq 1 \; \mbox{mod} \; p$ for any prime
divisor $q$ of $p-1$.  \\
\\ 
One choice for the constants would be $a=69069$, $c=0$ and $m = 2^{32}$.
Other choices are $a = 1 812 433 253$, $c=0$ and $m = 2^{32}$ or 
$a = 1 566 083 941$, $c=0$ and $m = 2^{32}$.
As a historical note, the first proposal was $a=23$, $c=0$ and $m=10^8+1$.
IBM used in the sixties and seventies the values
$a=65539$, $c=0$ and $m=2^{31}$ as well as $a=16807$, $c=0$ and $m=2^{31}-1$.
These generators are not recommended by today's standards.\\
\\
\bs
{\it Exercise \theexercise: Consider the simple multiplicative linear congruential
generator with $a=5$, $c=1$, $m=16$ and $s_0=1$. Write down the first twenty numbers
generated with this method. How long is the period ? Since $m=2^4$ write down the sequence
also in the binary representation and look at the lowest
order bits.
\stepcounter{exercise}}
\es
\\
\\
One of the drawbacks of the multiplicative linear congruential generator
was discovered by the spectral test. Here one considers the set of points
$(s_n, s_{n+1}, ..., s_{n+d-1})$ of consecutive pseudo-random numbers in $d$-dimensional
space. It was discovered that these points lie mainly in hyperplanes.\\
\\ 
\bs
{\it Exercise \theexercise: Generate a three- and a two-dimensional plot of points 
$(s_n,s_{n+1},s_{n+2})$ and $(s_n,s_{n+1})$ taken
from the generator $a=137$, $c=187$ and $m=256$. 
\stepcounter{exercise}}
\es
\\
\\
Since we work with a finite precision, even sequences from truly random numbers, truncated
to our precision, would reveal a lattice structure.
Let $1/\nu_1$ denote the distance between neighbouring points in a one-dimensional plot,
$1/\nu_2$ the distance between neighbouring lines in a two-dimensional plot,
$1/\nu_3$ the distance between neighbouring planes in a three-dimensional plot, etc.
The difference between truly random sequences, truncated to a certain precision, and sequences
from a multiplicative linear congruential generator is given by the fact, that in the former case
$1/\nu_d$ is indepenent of the dimension $d$, while in the latter case $1/\nu_d$ increases with
the dimension $d$.\\
\\ 
\bs
{\it Exercise \theexercise: 
We are interested in the integral
\bq
I & = & 2 \int\limits_0^1 dx \int\limits_0^1 dy \int\limits_0^1 dz \; \sin^2\left( 2 \pi (9x-6y+z) \right).
\eq
This integral can be solved analytically by doing the $z$-integration first and yields $1$.
Suppose we are ignorant about this possibility 
and perform a brute-force Monte Carlo integration using the
multiplicative linear congruential generator $a=65539$, $c=0$ and $m=2^{31}$.
Innocent as we are we use three consecutive random numbers from the generator to define a point
in the cube: $(x,y,z)_n = (s_{3n}/m, s_{3n+1}/m, s_{3n+2}/m)$.
What do you get ?
In order to see what went wrong show that three consecutive numbers of this generator
satisfy
\bq
\left( 9 s_n - 6 s_{n+1} + s_{n+2} \right) \; \mbox{mod} \; 2^{31} & = & 0.
\eq
\stepcounter{exercise}}
\es

\subsubsection{Lagged Fibonacci generator}

Each number is the result of an arithmetic operation
(usually addition, sometimes subtraction) between two numbers which have
occured somewhere earlier in the sequence, not necessarily the last two :
\begin{eqnarray}
s_{i} & = & ( s_{i-p} + s_{i-q} ) \;\mbox{mod}\; m.
\end{eqnarray}
A popular choice is:
\begin{eqnarray}
s_i & = & \left( s_{i-24} + s_{i-55} \right) \;\mbox{mod}\; 2^{32}.
\end{eqnarray}
It was proposed in 1958 by G.J. Mitchell and D.P. Moore. This generator has a period of
\begin{eqnarray}
2^f \left( 2^{55} - 1 \right),
\end{eqnarray}
where $0 \leq f < 32$.

\subsubsection{Shift register generator}

The generalized feedback shift register generator \cite{r250} is based on the
recurrence relation
\bq
s_i & = & s_{i-p} \oplus s_{i-p+q},
\eq
where $\oplus$ deontes the bitwise exclusive-or operation (
$0 \oplus 0 = 1 \oplus 1 = 0$, $0 \oplus 1 = 1 \oplus 0 = 1$).
One choice for lags is given by $p=250$ and $q=103$. With this choice
one has a period of $2^{250}-1$.

\subsubsection{RANMAR}

RANMAR \cite{ranmar} is combination of two generators. The first one is a lagged Fibonacci generator,
\bq
r_i & = & (r_{i-97} - r_{i-33}) \; \mbox{mod} \; 2^{24}.
\eq
The second part is a simple arithmetic sequence defined by
\bq
t_i & = & \left\{ \begin{array}{ll} 
t_{i-1} - 7654321, & \mbox{if} \; \; t_{i-1} - 7654321 \ge 0, \\
t_{i-1} - 7654321 + 2^{24} - 3 & \mbox{otherwise.} \\
\end{array} \right.
\eq
The final random number is then obtained as
\bq
s_i & = & (r_i - t_i) \; \mbox{mod} \; 2^{24}.
\eq 

\subsubsection{ACARRY/RCARRY/RANLUX}

Marsaglia, Narasimhan and Zaman \cite{acarry} proposed a lagged Fibonacci generator with a carry bit.
One first computes
\bq
\Delta_n & = & s_{n-i} - s_{n-j} - c_{n-1}, \;\;\;\;j>i\ge1,
\eq
and sets
\bq
\begin{array}{lll}
s_n = \Delta_n, & c_n=0, & \mbox{if} \;\; \Delta_n \ge 0, \\
s_n = \Delta_n + b, & c_n =1 & \mbox{otherwise.}\\
\end{array}
\eq
With the choice $j=24$, $i=10$ and $b=2^{24}$ this generator is known
under the name RCARRY.
One has observed that this algorithm fails the gap test and discovered
correlations between successive vectors of $j$ random numbers.
To improve the situation L\"uscher \cite{luescher} proposed to read $j$ numbers, and
to discard the following $p-j$ ones.
With the parameter $p=223$ the generator is known as RANLUX.

\subsection{Quasi-random numbers}

A major drawback of Monte Carlo integration with pseudo-random numbers is given by
the fact that the error term scales only as $1/\sqrt{N}$.
This is inherent to methods based on random numbers.
However in Monte Carlo integration the true randomness of the generated numbers
is not so much relevant. More important is to sample the integration region as uniform
as possible.
This leads to the idea to choose the points deterministically such that to minimize
the integration error.
If the integrand is sufficiently smooth one obtains a deterministic error bound,
which will scale like $N^{-1} \ln^p(N)$ for some $p$.\\
\\
We say a sequence of points is uniformly distributed if
\bq
\lim\limits_{N \rightarrow \infty} \frac{1}{N}
\sum\limits_{n=1}^N \chi_J(x_n) & = & \mbox{vol}(J)
\eq
for all subintervals $J \subset [0,1]^d$. Here $\mbox{vol}(J)$ denotes the
volume of $J$ and $\chi_J(x)$ the characteristic function of $J$, e.g.
$\chi_J(x) = 1$ if $x \in J$ and $\chi_J(x)=0$ otherwise.
As a measure of how much a finite sequence of points $x_1$, $x_2$, ..., $x_N$
deviates from the uniform distribution we define the discrepancy
\bq
\label{defdiscrepancy}
D & = & \sup\limits_{J \in {\cal J}} \left( 
\frac{1}{N} \sum\limits_{n=1}^N \chi_J(x_n) - \mbox{vol}(J) \right),
\eq
where the supremum is taken over a family ${\cal J}$ of
subintervals $J$ of $[0,1]^d$.
By specifying the family $\cal J$ we obtain two widely used concepts of discrepancy:
The extreme discrepancy is obtained from a family $\cal J$, where each subinterval $J$
contains the points $x=(u_1,...,u_d)$ with
\bq
u_i^{min} \le u_i < u_i^{max}, & & i=1,...,d.
\eq
A subinterval $J$ for the extreme discrepancy can therefore be specified
by giving the ``lower-left-corner'' $x^{min}=(u_1^{min},...,u_d^{min})$ and
the ``upper-right-corner'' $x^{max}=(u_1^{max},...,u_d^{max})$.
The star discrepancy is a special case of the extreme discrepancy
for which $u_1^{min} = ... = u_d^{min} = 0.$
The star discrepancy is often denoted by $D^\ast$.
One can show that
\bq
\label{discrepancy}
D^\ast \le D^{extreme} \le 2^d D^\ast.
\eq
\bs
{\it Exercise \theexercise: Prove this statement. \\
Hint: The first inequality is trivial. To prove the
second inequality you may start for $d=1$ from the fact 
that the number of points
in $[u_1^{min},u_1^{max}[$ equals the number of points in
$[0,u_1^{max}[$ minus the number in $[0,u_1^{min}[$.
\stepcounter{exercise}}
\es
\\
\\
If in the definition of the discrepancy (eq.~\ref{defdiscrepancy}) 
the supremum norm is replaced by $||...||^2$ one obtains the 
mean square discrepancy.\\
\\
Before coming to the main theorem for quasi-Monte Carlo integration, we first
have to introduce the variation $V(f)$ of the function $f(x)$ on the hypercube
$[0,1]^d$. We define
\bq
V(f) & = & \sum\limits_{k=1}^d \sum\limits_{1\le i_1<i_2<...<i_k\le d}
\left. V^{(k)}(f)\right|_{i_1,...,i_k},
\eq
where $V^{(k)}(\tilde{f})$ of a function $\tilde{f}(u_1,...,u_k)$
depending on $k$ variables $u_1$, ..., $u_k$ is given by
\bq
\label{subvariation}
V^{(k)}(f) & = & \int\limits_0^1 du_1 ... \int\limits_0^1 du_k
\left| \frac{\partial^kf(u_1,...,u_k)}{\partial u_1 ... \partial u_k} \right|
\eq
and $\left. V^{(k)}(f)\right|_{i_1,...,i_k}$ denotes
the restriction of $f$ to the $k$-dimensional face defined
by $(u_1,...,u_d) \in [0,1]^d$ and $u_j=1$ if $j\neq i_1,...,i_k$.
We further require that the partial derivatives in eq.~\ref{subvariation}
are continous.
If $V(f)$ is finite one says that $f$ is of bounded variation on $[0,1]^d$.\\
\\
We now have the main theorem of quasi-Monte Carlo integration: If $f$
has bounded variation on $[0,1]^d$ then for any $x_1,...,x_N \in [0,1]^d$
we have
\bq
\label{maintheoremquasiMC}
\left| \frac{1}{N} \sum\limits_{n=1}^N f(x_n) - \int dx f(x) \right|
\le V(f) D^\ast(x_1,...,x_N).
\eq
Eq.~\ref{maintheoremquasiMC} shows that the error will be bounded
by the product of the variation of the function times the 
discrepancy of the point set.
Since the variation of the function depends on the function $f$ alone,
but not on the chosen point set $x_1$,..., $x_N$, 
it is inherent to the problem.
The error may be reduced by choosing a point set with a low discrepancy.\\
\\
\bs
{\it Exercise \theexercise: Show that star discrepancy of the 
one-dimensional point
set $x_n=(2n-1)/(2N)$, $n=1,...,N$, is given by $D^\ast=1/(2N)$.
(It can be shown that this is the smallest discrepancy we can get in one
dimension with a finite set of $N$ points.
Note that the construction of the points $x_n$ depends on the predefined total
number of points $N$. 
It can be shown that there is no infinite sequence $x_1$, $x_2$, ..., whose
first $N$ elements have the theoretical minimal discrepancy $1/(2N)$.
Instead there is a theorem which states that for the discrepancy $D^\ast(N)$
of the first $N$ elements of an infinite sequence one has
$D^\ast(N) \ge cN^{-1}\ln(N)$ for infinitely many $N$.)
\stepcounter{exercise}}
\es
\\
\\
There is a general theorem which states that the star discrepancy of a
finite point set cannot be made smaller than a certain value, which
depends on the number of points $N$ and the dimensionality $d$.
In practical calculations one is not so much interested in a finite
sequence $x_1$,...,$x_N$ which attains the minimal discrepancy
for a given value of $N$, but more in an infinite sequence whose
first $N$ elements have a small discrepancy. Such a sequence
is called a low-discrepancy sequence.
The advantages are that one can change the value of $N$ in a 
quasi-Monte Carlo integration without loosing the previously
calculated function values.
Again there is a general theorem which states that the discrepancy
of such a sequence cannot decrease faster than $1/N$ for large $N$.
In practice one knows explicit sequences which decrease
like $N^{-1}\ln^p(N)$ for some $p$.
We will now turn our attention to the construction of such sequences.

\subsubsection{Richtmyer sequences}

One of the first quasi-random number generators which yielded an infinite
sequence in arbitrary dimension $d$ was given by
Richtmyer. The $n$-th point $x_n=(u_1,...,u_d)$ in the sequence
is given by
\bq
u_i & = & n S_i \; \mbox{mod} \; 1, \;\;\;\; i=1,...,d,
\eq
where the $S_i$ are constants which should be irrational numbers in order
to ensure an infinite period.
Since truly irrational numbers cannot be represented in computers
the standard choice is to take $S_i$ equal to the square root of the 
$i$-th prime number.

\subsubsection{Halton sequences}

Let $b\ge2$ be an integer number. Every integer $n\ge0$ has a unique
digit representation in base $b$,
\bq
n & = & \sum\limits_{j=0}^\infty a_j b^j,
\eq
where $a_j \in \{0,1,...,b-1\}$ and $a_j=0$ for sufficiently large $j$,
e.g. the sum is actually finite.
We then define
\bq
\phi_b(n) & = & \sum\limits_{j=0}^\infty a_j b^{-1-j}.
\eq
The sequence $\phi_b(1)$, $\phi_b(2)$, ..., is called the van der Corput
sequence in base $b$ and used as a building block for the Halton
sequence defined by
\bq
x_n & = & \left(\phi_{b_1}(n),...,\phi_{b_d}(n) \right).
\eq
If one uses the first $d$ prime numbers for the bases $b_1$, ..., $b_d$ one
can show the star discrepancy of the Halton sequence
satisfies for all $N\ge2$
\bq
D^\ast_{Halton}(N) & \le & A_d \frac{\ln^d(N)}{N} 
+ O\left( \frac{\ln^{d-1}(N)}{N} \right).
\eq
The coefficient $A_d$ depends only on the first $d$ prime numbers
$p_1$, ..., $p_d$ and is given by
\bq
A_d & = & \prod\limits_{k=1}^d \frac{p_k-1}{2\ln p_k}.
\eq
One can show that
\bq
\lim\limits_{d\rightarrow \infty} \frac{\ln A_d}{d \ln d} & = & 1,
\eq
e.g. $A_d$ grows stronger than exponentially as the number of dimensions
increases.
This is the major drawback of the Halton sequence and limits
its applications in quasi-Monte Carlo integration to low dimensions.

\subsubsection{Sobol sequences}

Sobol sequences \cite{sobol} are obtained by first choosing a primitive polynomial over 
${\mathbb Z}_2$ of 
degree $g$:
\bq
P & = & x^g + a_1 x^{g-1} + ... + a_{g-1} x + 1,
\eq
where each $a_i$ is either $0$ or $1$.
The coefficients $a_i$ are used in the recurrence relation
\bq
v_i & = & a_1 v_{i-1} \oplus a_2 v_{i-2} \oplus ... 
\oplus a_{g-1} v_{i-g+1} \oplus v_{i-g} \oplus [v_{i-g}/2^g],
\eq
where $\oplus$ denotes the bitwise exclusive-or operation and each $v_i$ is a number which can be written
as $v_i = m_i/(2^i)$ with $0<m_i<2^i$.
Consequetive quasi-random numbers are then obtained from the relation
\bq
x_{n+1} & = & x_n \oplus v_c,
\eq
where the index $c$ is equal to the place of the rightmost zero-bit in the
binary representation of $n$. For example $n=11$ has the binary
representation $1011$ and the rightmost zero-bit is the third one.
Therefore $c=3$ in this case.
For a primitive polynomial of degree $g$ we also have to choose the first
$g$ values for the $m_i$. The only restriction is that the $m_i$ are all
odd and $m_i < 2^i$.
There is an alternative way of computing $x_n$:
\bq
x_{n} & = & g_1 v_1 \oplus g_2 v_2 \oplus ...
\eq
where $...g_3 g_2 g_1$ is the Gray code representation of $n$. 
The basic property of the Gray code is that the representations for $n$ and $n+1$
differ in only one position.
The Gray code representation  can be obtained from the binary representation according to
\bq
\label{graycode}
... g_3 g_2 g_1 & = & ... b_3 b_2 b_1 \oplus ... b_4 b_3 b_2.
\eq
\bs
{\it Exercise \theexercise: Find the Gray code representation for $n=0,1,...,7$.\\
Hint: Start from the binary representation and use the formula eq.~\ref{graycode}.
You should find $0$, $1$, $11$, $10$, $110$, $111$, $101$, $100$.
\stepcounter{exercise}}
\es \\
\\
The star discrepancy of the Sobol sequence satisfies
\bq
D^\ast & \le & B_d \frac{\ln^d N}{N} + O\left( \frac{\ln^{d-1}N}{N} \right),
\eq
where 
\bq
B_d & = & \frac{2^{\tau_d}}{d! (\ln2)^d} 
\eq
and
\bq
k \frac{d \ln d}{\ln \ln d} \le \tau_d \le \frac{d \ln d}{\ln 2} + \frac{d \ln \ln d}{\ln 2}
+ o( d \ln \ln d)
\eq
and asymptotically one finds $\ln B_d = O(d \ln \ln d)$ and $B_d$ increases also stronger
than exponentially with $d$.

\subsubsection{Faure sequences}

Let $d\ge3$ and let $b$ be the first prime number greater or equal to $d$.
We describe here the construction of the Faure sequence $x_1$, $x_2$, ... \cite{faure}.
Let $u_i$ be the components of the $n$-th point in the sequence:
$x_n=(u_1,u_2,...,u_d)$.
One starts from the digit representation
of $n-1$ in the base $b$:
\bq
n-1 & = & \sum\limits_{r=0}^\infty a_j b^j.
\eq
$u_1$ is given by
\bq
u_1 & = & \phi_b(n-1) = \sum\limits_{j=0}^\infty a_j b^{-1-j}
\eq
and $u_2$, $u_3$, ..., $u_d$ are obtained recursively from
\bq
u_{i+1} = C(u_i), & & 1 \le i < d.
\eq
If $v=\sum v_j b^{-1-j}$ and $w=\sum w_j b^{-1-j}$ then the transformation
$w = C(v)$
is defined by
\bq
w_j & = & \sum\limits_{i \ge j} \left( \begin{array}{c} i \\ j \\ \end{array} \right)
v_i \; \mbox{mod} \; b.
\eq
It can be shown that the star discrepancy of such a sequence satisfies
\bq
D^\ast & \le & C_d \frac{\ln^d N}{N} + O\left( \frac{\ln^{d-1}N}{N} \right)
\eq
where 
\bq
C_d & = & \frac{1}{d!} \left( \frac{b-1}{2 \ln b} \right)^d.
\eq
Asymptotically $C_d$ goes to zero.

\subsubsection{Niederreiter sequences}

Let $d\ge3$ and let $b$ be a prime power greater or equal to $d$.
As before we use the digit representation for $n$ in the base $b$
\bq
n & = & \sum\limits_{r=0}^\infty a_r b^r.
\eq
The $n$-th element $x_n=(u_1,...,u_d)$ of the Niederreiter sequence \cite{Niederreiter}
is given by
\bq
u_i & = & \sum\limits_{j=1}^\infty b^{-j} \left( \sum\limits_{r=j-1}^\infty
\left( \begin{array}{c} r \\ j-1 \\ \end{array} \right)
a_r c_i^{r-j+1} \; \mbox{mod} \; b \right),
\eq
where the $c_i$ are distinct elements from the set $0,1,...,b-1$.
Note that the sums are actually finite.
It can be shown that the star discrepancy of such a sequence satisfies
\bq
D^\ast & \le & D_d \frac{\ln^d N}{N} + O\left( \frac{\ln^{d-1}N}{N} \right),
\eq
where 
\bq
D_d & = & \frac{1}{d!} \frac{b-1}{2[b/2]} \left( \frac{[b/2]}{\ln b}
\right)^d.
\eq
Further
\bq
\lim\limits_{d\rightarrow \infty} \frac{\ln D_d}{d \ln \ln d} & \le & -1
\eq
and the coefficients $D_d$ decrease stronger than exponentially in the limit
$d \rightarrow \infty$.

\subsection{Summary on random numbers}

Monte Carlo integration relies on pseudo-random number generators. We reviewed some of the most
widely used algorithms for the generation of random numbers. 
Pseudo-random number generators might introduce correlations between succesive generated
numbers.
It is therefore recommended to check a Monte Carlo integration by repeating the same calculation
with a different generator.\\
\\
Quasi-random numbers are deterministic and designed to sample a $d$-dimensional
space as uniform as possible.
If the integrand is sufficiently smooth, the error bound is deterministic and 
will decrease like 
\bq
A \frac{\ln^d N}{N} + O \left( \frac{\ln^{d-1} N}{N} \right).
\eq
If the integrand has discontinuities 
(for example due to complicated integration boundaries, which have to be embedded into a hypercube)
the theorem on the error bound
is no longer valid and one usually estimates the error term like in Monte Carlo integration
with pseudo-random numbers.
The coefficient $A$ of the leading term of the error estimate depends on the dimension $d$ and goes 
for $d\rightarrow \infty$ to
infinity for the Halton and Sobol sequences, and to zero
for the Faure and Niederreiter sequences. Very little is known about the subleading terms.
Explicit simualtions with roughly $N=10^5$ points by F. James, J. Hoogland and R. Kleiss \cite{hoogland} show
that the quadratic discrepancy of these sequences is better than the one from pseudo-random numbers
provided $d \lesssim 12$, and approaches the latter one in higher dimensions.\\
\\
Further reading: The book of Knuth \cite{ran3} contains an excellent introduction to 
pseudo-random numbers, further I also used the review article by James \cite{ran2} in the preparation of this
section.
The article by I. Vattulainen et al. \cite{vattulainen} contains a comparison of various pseudo-random
number generators in several tests.\\
The section on quasi-random numbers is mainly based on the book by Niederreiter \cite{Niederreiter}.
Fox and Bratley and Fox \cite{fox} give a numerical implementation of several quasi-random number
generators.

\section{Generating samples according to a specified distribution}

Quite often one needs random numbers which are distributed according to a 
specified probability density function $p(x)$.
We will denote the cumulative distribution function by
\bq
P(x_{max}) & = & \int\limits_{-\infty}^{x_{max}} p(x) dx,
\eq
with appropriate generalizations in higher dimensions.
$P(x_{max})$ gives the probability that $x \le x_{max}$.
In the previous section we have seen how to generate pseudo-random or 
quasi-random numbers which are uniformly distributed in the interval
$[0,1]$.
The problem can therefore be specified as follows: Given a sequence of
random numbers, uniformly distributed in $[0,1]$, find a transformation such
that the resulting sequence is distributed according to $p(x)$.

\subsection{General algorithms}

\subsubsection{The inverse transform method}

We describe the inverse transform method in one dimension. Let $x$ be a random
variable distributed with density $p(x)$. The distribution function $P(x)$
takes values in the interval $[0,1]$. Let $u$ be a random variable uniformly
distributed in $[0,1]$. We set
\bq
x & = & P^{-1}(u).
\eq
For the differentials we have
\bq
p(x) dx & = & du.
\eq
In order to use the inverse transform method we have to know the function
$P^{-1}(u)$. This will not always be the case.\\
\\
One application of the inverse transform method is importance sampling.

\subsubsection{Acceptance-rejection method}

The acceptance-rejection method, developed by von Neumann, can be used
when an analytic form of $P(x)$ is not known. We assume that we can
enclose $p(x)$ inside a shape which is $C$ times an easily generated
distribution $h(x)$. Very often one chooses $h(x)$ to be a uniform 
distribution or a normalized sum of uniform distributions. Since
$p(x) \le C h(x)$ and both $p(x)$ and $h(x)$ are normalized to unity, one
has $C \ge 1$.
One first generates $x$ according to the distribution $h(x)$ and calculates
then $p(x)$ and $C h(x)$. Secondly, one generates a random number $u$,
uniformly distributed in $[0,1]$, and checks $u C h(X) \le p(x)$. If this
is the case, one accepts $x$, otherwise one rejects $x$ and starts again.
The efficiency of this method is $1/C$, therefore one tries to choose
$h(x)$ to be as close to $p(x)$ in shape as possible. 

\subsubsection{Applications}
\label{application}

One often encounters the situation that one needs random variables
distributed according to a Gaussian distribution:
\bq
p(x) & = & \frac{1}{\sqrt{2\pi \sigma^2}} e^{-\frac{(x-\mu)^2}{2 \sigma^2}}.
\eq
The Box-Muller algorithm gives a prescription to generate two independent
variables $x_1$ and $x_2$, distributed according to a Gaussian 
distribution with mean $\mu=0$ and variation $\sigma^2=1$ from two
independent variables $u_1$ and $u_2$, uniformly distributed in
$[0,1]$:
\bq
x_1 & = & \sqrt{-2 \ln u_1} \cos(2 \pi u_2), \nonumber \\
x_2 & = & \sqrt{-2 \ln u_1} \sin(2 \pi u_2).
\eq
\bs
{\it Exercise \theexercise: Prove this statement. \\
Hint: You should show that
\bq
du_1 du_2 & = & \frac{1}{2\pi} e^{-\frac{x_1^2+x_2^2}{2}} dx_1 dx_2
\eq
and that $x_1$ and $x_2$ are independent given the fact that $u_1$ and $u_2$
are independent.
\stepcounter{exercise}}
\es
\\
\\
Algoritms for generating many different distributions are known and can
be found in \cite{distribution}. We have collected some
``cooking recipes'' in the appendix B.

\subsection{The Metropolis algorithm}

In practical application one often wants to generate random variables
according to some probability density $p(x_1,...,x_d)$, which not
necesarrily factorizes. The methods described so far are in most
cases insufficient. 
In practice one often uses the Metropolis algorithm \cite{metropolis} 
for generating random samples
which are distributed according to a multivariate probability density
$p(x_1,...,x_d)$  where $d$ is large.
Let us call the vector $\phi=(x_1,...,x_d)$ a state of the ensemble, which
we want to generate.
Within the Metropolis algorithm one starts from a state $\phi_0$, and
replaces iteratively an old state by a new one, in such a way,
that the correct probability density distribution is obtained in the limit
of a large number of such iterations.
The equilibrium distribution is reached, regardless of the state one started
with. Once the equilibrium distribution is reached, repeated application
of the algorithm keeps one in the same ensemble.
In short, the desired distribution is the unique fix point of the algorithm.
Two important conditions have to be met for the Metropolis algorithm to work:
Ergodicity and detailed balance.
Detailed balance states that the transition probabilities 
$W(\phi_1 \rightarrow \phi_2)$ and $W(\phi_2 \rightarrow \phi_1)$
obey
\bq
p(\phi_1) W(\phi_1 \rightarrow \phi_2) & = & p(\phi_2) W(\phi_2 \rightarrow \phi_1).
\eq
Ergodicity requires that each state can be reached from any other state
within a finite number of steps. 
Given a state $\phi_1$, one iteration of the Metropolis algoritm
consists of the following steps:
\begin{enumerate}
\item Generate (randomly) a new candidate $\phi'$.
\item Calculate $\Delta S = - \ln(p(\phi')/p(\phi_1))$.
\item If $\Delta S < 0$ set the new state $\phi_2=\phi'$.
\item If $\Delta S > 0$ accept the new candidate only with
probability $p(\phi')/p(\phi)$, otherwise retain the old state
$\phi_1$.
\item Do the next iteration.
\end{enumerate}
Step 3 and 4 can be summarized that the probability of accepting the 
candidate $\phi'$ is given by 
$W(\phi_1 \rightarrow \phi')=\mbox{min}(1,e^{-\Delta S})$.
It can be verified that this transition probabilty satisfies detailed
balance.
The way how a new candidate $\phi'$ suggested is arbitrary, 
restricted only by the condition that one has to be able to reach each
state within a finite number of steps.\\
\\
The Metropolis algorithm also has several drawbacks: Since one can start from
an arbitrary state it takes a number $\tau_t$ of steps to reach the desired
equilibrium distribution. 
It can also be possible that one reaches a metastable state, after which one
would get out only after a large (and generally unknown) number of steps.
One therefore starts the Metropolis algorithm with several different
initializations and monitors if one reaches the same equilibrium state.
Furthermore, once the equilibrium is reached,
succesive states are in general highly correlated. 
If one is interested
in an unbiased sample of states $\phi_i$, one therefore generally discards
a certain number $\tau_d$ of states produced by the Metropolis algorithm, before
picking out the next one. The number of time steps $\tau_d$ is called the decorrelation
time and is of order $\tau_d = \xi^2$, where $\xi$ is a typical correlation lenght
of the system. 
This is due to the fact that the Metropolis algorithm updates the states locally at random.
One therefore 
performs a random walk through configuration space, and it takes therefore
$\xi^2$ steps to move a distance $\xi$ ahead. In certain applications for
critical phenomena the correlation lenght $\xi$ becomes large, which makes it
difficult to obtain unbiased samples.\\
\\
Since new candidates in the Metropolis algorithm are picked out randomly, the
``random-walk-problem'' is the origin of inefficiencies to obtain the equilibrium distribution
or to sample uncorrelated events.
One way to improve the situation is to use a-priori probabilities \cite{krauth}.
The transistion probability is written as
\bq
W(\phi_1 \rightarrow \phi_2) & = & A(\phi_1 \rightarrow \phi_2) \tilde{W}(\phi_1 \rightarrow \phi_2).
\eq
where $A(\phi_1 \rightarrow \phi_2)$ is the a-priori probability that state $\phi_2$ is suggested,
given the fact that the system is in state $\phi_1$.
The probability of accepting the new candidate is now given by
\bq
\mbox{min}\;\left( 1, \; \frac{A(\phi_2 \rightarrow \phi_1)}{A(\phi_1 \rightarrow \phi_2)}
\frac{p(\phi_2)}{p(\phi_1)} \right).
\eq
The bias we introduced in suggesting new candidates is now corrected by the acceptance rate.
By choosing the a-priori probabilities carefully, one can decrease the rejection rate and improve
the efficiency of the simulation.
As an example consider the case where a state $\phi=(x_1,...,x_d)$
describes a dynamical system, like a molecular gas. In this case a discretized or
approximative version of the equations of motion can be used
to suggest new candidates and the Metropolis algorithm might converge
faster. The combination of molecular dynamics (e.g. the use of the equations of motion)
with Monte Carlo methods is sometimes
called hybrid Monte Carlo.

\subsubsection{Numerical simulations of spin glasses}

The Ising and the Potts model are widely used in statistical physics.
Both describe an ensemble of spins interacting with each other through
next-neighbour interactions.
The Hamiltonian of the Ising model
is given by
\bq
H_{Ising} & = & - J \sum\limits_{\l i,j \r} S_i S_j,
\eq
where $J>0$ is a constant, $\l i,j \r$ denotes the sum over all next-neighbours
and the spin $S_i$ at site $i$ takes the values $\pm 1$.
The Hamiltonian of the Potts model is given by
\bq
H_{Potts} & = & - \sum\limits_{\l i,j \r} J_{ij} \left( \delta_{S_i S_j} - 1 \right).
\eq
Here each spin $S_i$ may take the values $1$, $2$, ..., $q$ and $\delta_{ab}$ denotes the
Kronecker delta. We have also allowed the
possibility of different couplings $J_{ij}$ depending on the sites $i$ and $j$.
Let $\phi=(S_1, S_2, ..., S_N)$ denote a state of the model. Observables are given
by
\bq
\l O \r & = & \frac{\sum O(\phi) e^{-H(\phi)/kT}}{\sum e^{-H(\phi)/kT}},
\eq
where the sum runs over all possible states. One observable is the magnetization $M$
or the order parameter. For the Ising model the magnetization is given by
\bq
M_{Ising} & = & \frac{1}{N} \sum S_i
\eq
and for the Potts model by
\bq
M_{Potts} & = & \frac{q \; \mbox{max} ( N_\alpha ) - 1 }{q-1},
\eq
where $N_\alpha$ is the number of spins with value $\alpha$.
A second observable is the magnetic susceptibility given by
\bq
\chi & = & \frac{N}{k T} \left( \l M^2 \r - \l M \r^2 \right).
\eq
\bs
{\it Exercise \theexercise: Write a Monte Carlo program which simulates the
two-dimensional Ising model on a $ 16 \times 16$ lattice with periodic
boundary conditions using the Metropolis algorithm.
(Note that with a $16 \times 16$ lattice, the total number of states is $2^{256}$. 
This is far too large to evaluate the partition sum by brute force.)
You may initialize the model with all spins up (cold start) or with a random
distribution of spins (hot start).
Step through the 256 spins one at a time making an attempt to flip the current spin.
Plot the absolute value of the magnetization for various values of $K=J/kT$ between
$0$ and $1$. Is anything happening around $K=0.44$ ?
\stepcounter{exercise}}
\es\\
\\
To simulate the Potts model it is more efficient to use the Swendsen-Wang algorithm \cite{wang}
which can flip clusters of spins in one step, instead of the standard Metropolis algorithm, which 
has to build up a spin flip of a cluster through succesive steps.
The Swendsen-Wang algorithm introduces for each neighbouring pair $i$ and $j$ a bond $n_{ij}$ between
the two spins. The bond variable $n_{ij}$ can take the values $0$ and $1$, the last value signifies
that the two spins are ``frozen'' together.
One iteration of the Swendsen-Wang algorithm works as follows:
\begin{enumerate}
\item For each neighbouring pair $i$, $j$ the spins are frozen with probability
\bq
P_{ij} & = & 1 - \exp \left( - \frac{J_{ij}}{kT} \delta_{S_i S_j} \right).
\eq
\item Having gone over all pairs, clusters are now formed in a second step:
A cluster contains all spins which have a path of frozen bonds connecting them.
Note that by constrution all spins in one cluster have the same value.
\item For each cluster a new value for the cluster spin is chosen randomly
and with equal probability.
\end{enumerate}
The reason why this algorithm works is roughly as follows:
The Swendsen-Wang method enlarges the set of variables from the $S_i$ to the set
$S_i$ and $n_{ij}$. Further, the partition function of the Potts model can be rewritten
as
\bq
Z & = & \sum\limits_{ \{ S_i \} } \exp(-H/kT) \nonumber \\
& = & \sum\limits_{ \{ S_i \} } \sum\limits_{ \{ n_{ij} \} } \prod\limits_{ \l i,j \r}
\left( (1-P_{ij}) \delta_{0,n_{ij}} + P_{ij} \delta_{1,n_{ij}} \delta_{S_i S_j}
\right).
\eq
Integrating out the $n_{ij}$ (e.g. marginalization with respect to the bonds $n_{ij}$) 
one recovers the Potts model. 

\subsubsection{Numerical simulations of quantum field theories}

Observables in quantum field theories are given by
\bq
\l O \r & = & \frac{\int {\cal D} \phi \; O(\phi) e^{-S(\phi)} }
{\int {\cal D} \phi \; e^{-S(\phi)}},
\eq
where $S(\phi)$ is the Euclidean action of the field $\phi$. The
basic idea of lattice field theory is to approximate the infinite
dimensional path integral by a finite sum of field configurations $\phi_i$.
Since the factor $e^{-S(\phi)}$ can vary over several order of magnitudes,
simple random sampling will yields poor results, and one therefore uses
importance sampling and chooses $\phi$
according to some propability $P(\phi)$:
\bq
\l O \r & \approx & 
\frac{ \sum\limits_{i=1}^N O(\phi_i) P^{-1}(\phi_i) e^{-S(\phi_i)} } 
{ \sum\limits_{i=1}^N P^{-1}(\phi_i) e^{-S(\phi_i)} }.
\eq
The action $S(\phi_i)$ is also approximated by a finite sum. 
For example the discretized version of the action of $\phi^4$-theory
\bq
S & = & \int d^dx \left( \frac{1}{2} (\partial_\mu \phi) (\partial^\mu \phi)
+ \frac{1}{2} m^2 \phi^2 + \frac{g}{4!} \phi^4 \right)
\eq
is given in dimensionless quantities by
\bq
S(\phi_i) & = & \sum\limits_x \left(
- \kappa \left( \sum\limits_{\mu=0}^{d-1} \phi(x)
\left( \phi(x+a \hat{\mu}) + \phi(x-a \hat{\mu}) \right) \right)
+ \phi(x)^2 + \lambda \phi(x)^4 \right),
\eq
where $a$ is the lattice spacing and  $\hat{\mu}$ is a unit vector in the direction
$\mu$. 
The field has been rescaled according to
\bq
\phi_{cont} & = & \sqrt{2 \kappa} a^{1-d/2} \phi_{discr}
\eq
The parameters $\kappa$ and $\lambda$ are related 
up to correction of order $a^2$ to the original parameters
$m^2$ and $g$ by
\bq
m^2 a^2 = \frac{1}{\kappa}-2d, & & 
g a^{4-d} = \frac{6 \lambda}{\kappa^2}.
\eq
The perfect choice for $P(\phi_i)$ would be $P_{eq}(\phi_i) \cong e^{-S(\phi_i)}$,
but this would require that we know the partition function analytically.
One is therefore forced to construct a random walk through configuration
space such that
\bq
\lim\limits_{n \rightarrow \infty} P(\phi) & = & P_{eq}(\phi).
\eq
This is usually done with the help of the Metropolis algorithm. One first
chooses a random change in the field configuration 
$\phi_i \rightarrow \phi_i'$, calculates the change in the action
$\Delta S = S(\phi_i')-S(\phi)$ and the Metropolis transition
probabilty  $W=\mbox{min}(1,\exp(-\Delta S))$. One then throws a random
number $u$ uniformly distributed in $[0,1]$ and accepts the new
configuration $\phi_i'$ if $u<W$, otherwise $\phi_i'$ is rejected.

\subsection{Generating phase space for particle collisions}

Observables in high-energy collider experiments are often of the following
form
\bq
O & = & \int d\Phi_n(p_a+p_b,p_1,...,p_n) \; \frac{{\cal M} }{8 K(s)} \; \Theta(O,p_1,...,p_n),
\eq
where $p_a$ and $p_b$ are the momenta of the incoming particles, the outgoing
particles are denoted by the labels $1$,...,$n$. The Lorentz-invariant phase space
is denoted by $d\Phi_n$, $1/(8K(s))$ is a kinematical factor with $s=(p_a+p_b)^2$ which
includes the averaging over initial spins (we assumed two spin states for each initial
particle), $\cal M$ is the relevant matrix element squared and $\Theta(O,p_1,...,p_n)$
is a function which defines the observable and includes all experimental cuts.
The situation described above corresponds to electron-positron annihilation. If
hadrons appear in the initial state there are slight modifications.
The exact matrix element squared is often impossible to calculate.
Event generators like HERWIG \cite{herwig} or PYTHIA \cite{pythia} approximate
the matrix element squared through a three-stage process: First there is a perturbative
hard matrix element for the hard subprocess, where only a few partons are involved.
Secondly, the partons originating from the hard scattering process are then allowed to radiate
off additional partons. This stage is usually called parton showering and results in a cascade
of partons. Finally the resulting partons are converted into observable hadrons with the
help of a phenomenological model.
Perturbative calculations follow a different philosophy: First, they restrict the set of observables
to ones which are infrared safe, e.g. which can reliable be calculated in perturbation theory.
Secondly, they are based on parton-hadron duality and calculate an observable in the parton picture.
This amounts to say, that one assumes that hadronization corrections are small.
The matrix element squared $\cal M$ is then calculated order by order in perturbation
theory. The leading order term can be more or less obtained from automized procedures,
the standard of today is a next-to-leading order calculation, the frontier of tomorrow are 
next-to-next-to-leading order calculations.
Bot in perturbative calculations and in event generators the problem arises to generate
the four-momenta of the outgoing particles.
The Lorentz-invariant phase space $d \Phi_n$ for $n$ particles with momenta $p_1$, ..., $p_{n}$ 
and masses $m_1$, ..., $m_n$ is given by
\begin{eqnarray}
d \Phi_n(P,p_1,..,p_n) & = & 
\prod\limits_{i=1}^n \frac{d^4 p_i}{(2 \pi)^{3}} \Theta(p_i^0) \delta(p_i^2-m_i^2) 
(2 \pi)^4 \delta^4\left( P - \sum\limits_{i=1}^n p_i \right)\nonumber \\
& = & 
\prod\limits_{i=1}^n \frac{d^{3} p_i}{(2 \pi)^{3} 2 E_i} 
(2 \pi)^4 \delta^4\left( P - \sum\limits_{i=1}^n p_i \right).
\end{eqnarray}
The phase space volume for massless particles $m_1=...=m_n=0$ is
\begin{eqnarray}
\Phi_n & = & \int d \Phi_n = (2 \pi)^{4-3n} \left( \frac{\pi}{2} \right)^{n-1} \frac{(P^2)^{n-2}}{\Gamma(n) \Gamma(n-1)} .
\end{eqnarray}
The phase space factorizes according to
\begin{eqnarray}
\label{phasespacefactorization}
d \Phi_n(P,p_1,...,p_n) & = & 
\frac{1}{2 \pi} dQ^2 d\Phi_j(Q,p_1,...,p_j) d \Phi_{n-j+1}(P,Q,p_{j+1},...,p_n),
\end{eqnarray}
where $Q = \sum\limits_{i+1}^j p_i$.

\subsubsection{Sequential approach}

One possibility to generate the $n$-particle phase space is to consider sequentially
two-body decays \cite{kajantie}. Using eq.~\ref{phasespacefactorization} one arrives at
\bq
d\Phi_n & = & \frac{1}{(2 \pi)^{n-2}} dM_{n-1}^2 ... dM_2^2 d\Phi_2(n) ... d\Phi_2(2)
\eq
with $M_i^2=q_i^2$, $q_i = \sum\limits_{j=1}^i p_i$ and $d\Phi_2(i) = d\Phi_2(q_i,q_{i-1},p_i)$.
The allowed region for the invariant masses $M_i$ is given by
$(m_1+...+m_i)^2 \le M_i^2 \le (M_{i+1} - m_{i+1})^2$. 
In the rest frame of $q_i$ the two-particle phase space $d\Phi_2(q_i,q_{i-1},p_i)$ is given
by
\bq
d\Phi_2(q_i,q_{i-1},p_i) & = & \frac{1}{(2\pi)^2} \frac{\sqrt{\lambda(q_i^2,q_{i-1}^2,m_i^2)}}{8 q_i^2}
d \varphi_i d (\cos \theta_i),
\eq
where the triangle function $\lambda(x,y,z)$ is defined by
\bq
\lambda(x,y,z) & = & x^2 + y^2 + z^2 - 2xy - 2yz - 2 zx.
\eq
This suggests the following algorithm:
\begin{enumerate}
\item Set $i=n$, $q_i=P$ and $M_i=\sqrt{q_i^2}$.
\item Transform to the rest frame of $q_i$.
\item Use two random variables $u_{i1}$, $u_{i2}$ and set $\varphi_i=2\pi u_{i1}$,
$\cos \theta_i = u_{i2}$.
\item If $i \ge 3$ use a third random variable $u_{i3}$ and set
$M_{i-1}= (m_1+...+m_{i-1})+u_{i3}(M_{i}-m_i)$, 
otherwise if $i =2$ set $M_1=m_1$.
\item Set 
\bq
|{\vec{p}_i}\;'| & = & \frac{\sqrt{\lambda(M_i^2,M_{i-1}^2,m_i^2)}}{2 M_i}
\eq
and
${\vec{p}_i}\;'=|{\vec{p}_i}\;'| \cdot (\sin \theta_i \sin \varphi_i,
\sin \theta_i \cos \varphi_i, \cos \theta_i)$.
Set further
\bq
p_i' = (\sqrt{|{\vec{p}_i}\;'|^2+m_i^2},{\vec{p}_i}\;'), & &
q_{i-1}' = (\sqrt{|{\vec{p}_i}\;'|^2+M_{i-1}^2}, -{\vec{p}_i}\;').
\eq
\item Transform back to the original Lorentz system.
\item Decrease $i=i-1$. If $i\ge2$ go back to step 2, otherwise set $p_1=q_1$.
\end{enumerate}
The weight of the generated event depends on the event and is given by
\bq
w & = & (2\pi)^{4-3n} 2^{1-2n} \frac{1}{M_n} \prod\limits_{i=2}^n 
\frac{\sqrt{\lambda(M_i^2,M_{i-1}^2,m_i^2)}}{M_i}.
\eq
\bs
{\it Exercise \theexercise: Lorentz transformation. Let $q$ be a timelike fourvector with $q^2=M^2$.
Denote by $S$ the coordinate system in which $q$ is given, and by $S'$ the rest frame of $q$, e.g.
in $S'$ we have $q'=(M,0,0,0)$. Let $p$ be an arbitrary fourvector in $S$ and denote
the corresponding representation in $S'$ by $p'$. Show that $p$ and $p'$ are related
by
\bq
p_t = \gamma \left( p_t' + \frac{\vec{p}\;' \vec{q}}{q_t} \right), & & 
\vec{p} = \vec{p}\;' + \vec{q} \left( (\gamma -1 ) \frac{\vec{p}\;' \vec{q}}{|\vec{q}|^2}
+ \gamma \frac{p_t'}{q_t} \right),
\eq
where $\gamma=q_t/M$.
Show also that the inverse transformation is given by
\bq
p_t' = \gamma \left( p_t - \frac{\vec{p} \vec{q}}{q_t} \right), & & 
\vec{p}\;' = \vec{p} + \vec{q} \left( (\gamma -1 ) \frac{\vec{p} \vec{q}}{|\vec{q}|^2}
- \gamma \frac{p_t}{q_t} \right).
\eq
\stepcounter{exercise}}
\es

\subsubsection{Democratic approach}

Where as the sequential algorithm discussed above generates an event with a weight depending
on the generated fourvectors, it is sometimes desirable to have an algorithm which
sweeps out phase space (almost) uniformly.
The RAMBO-algorithm \cite{ran6} maps a hypercube $[0,1]^{4n}$ of random numbers into $n$ 
physical four-momenta with center-of-mass energy $\sqrt{P^2}$.
Massless fourvectors can be generated with uniform weight, and we discuss this case first.\\
\\
Let $P=(P,0,0,0)$ be a time-like four-vector. 
The phase space volume for a system of $n$ massless particles with
center-of-mass energy $\sqrt{P^2}$ is
\begin{eqnarray}
\label{phase_space_measure}
\Phi_n & = & \int \prod\limits_{i=1}^n \frac{d^4 p_i}{(2 \pi)^3} \theta(p_i^0) \delta(p_i^2)  
(2 \pi)^4 \delta^4\left( P - \sum\limits_{i=1}^n p_i \right).
\end{eqnarray}
To derive the RAMBO-algorithm one starts instead from the quantity
\begin{eqnarray}
R_n & = & \int \prod\limits_{i=1}^n \frac{d^4 q_i}{(2 \pi)^3} \theta(q_i^0) \delta(q_i^2)  
(2 \pi)^4 f(q_i^0) 
= (2 \pi)^{4-2n} \left( \int\limits_0^\infty x f(x) dx \right)^n.
\end{eqnarray}
The quantity $R_n$ can be interpreted as describing a system of $n$ massless four-momenta $q_i^\mu$
that are not constrained by momentum conservation but occur with some weight function $f$ which
keeps the total volume finite.
The four-vectors $q_i^\mu$ are then related to the physical four-momenta $p_i^\mu$ by the
following Lorentz and scaling transformations:
\begin{eqnarray}
\label{boost_and_scale}
p_i^0 = x \left( \gamma q_i^0 + \vec{b} \cdot \vec{q}_i \right), & &
\vec{p}_i = x \left( \vec{q}_i + \vec{b} q_i^0 + a \left( \vec{b} \cdot \vec{q}_i \right) \vec{b} \right),
\end{eqnarray}
where
\begin{eqnarray}
& & Q^\mu = \sum\limits_{i=1}^n q_i^\mu, \;\;\; M = \sqrt{Q^2}, \;\;\; 
\vec{b} = - \frac{1}{M} \vec{Q}, \nonumber \\
& & \gamma = \frac{Q^0}{M} = \sqrt{1 + \vec{b}^2}, \;\;\; a = \frac{1}{1+\gamma}, \;\;\; x = \frac{\sqrt{P^2}}{M}.
\end{eqnarray}
Denote this transformation and its inverse as follows
\begin{eqnarray}
p_i^\mu = x H^\mu_{\vec{b}}(q_i), & & q_i^\mu = \frac{1}{x} H^\mu_{-\vec{b}}(p_i).
\end{eqnarray}
By a change of variables one can show
\begin{eqnarray}
R_n & = & \int \prod\limits_{i=1}^n \left( \frac{d^4 p_i}{(2 \pi)^3} \delta(p_i^2) \theta(p_i^0) \right)
(2 \pi)^4 \delta^4 \left(P - \sum\limits_{i=1}^n p_i \right) \nonumber \\
& &\cdot \left( \prod\limits_{i=1}^n f \left( \frac{1}{x} H^0_{-\vec{b}}(p_i) \right) \right)
\frac{(P^2)^2}{x^{2n+1} \gamma} d^3b dx.
\end{eqnarray}
With the choice $f(x) = e^{-x}$ the integrals over $\vec{b}$ and $x$ may be performed and 
one obtains
\begin{eqnarray}
R_n & = & \Phi_n \cdot S_n
\end{eqnarray}
with 
\begin{eqnarray}
S_n & = & 2 \pi (P^2)^{2-n} \frac{ \Gamma\left(\frac{3}{2}\right) \Gamma(n-1) \Gamma(2n)}{\Gamma\left(n+\frac{1}{2}\right)}.
\end{eqnarray}
This gives a Monte Carlo algorithm which generates massless four-momenta $p_i^\mu$ according to the phase-space
measure (\ref{phase_space_measure}).
The algorithm consists of two steps:
\begin{enumerate}
\item Generate independently $n$ massless four-momenta $q_i^\mu$ with isotropic angular distribution and 
energies $q_i^0$ distributed according to the density $q_i^0 e^{-q_i} dq_i^0$.
Using $4n$ random numbers $u_i$ uniformly distributed in $\left[0,1\right]$ this is done as follows:
\begin{eqnarray}
c_i = 2 u_{i_1} -1, & \varphi_i = 2 \pi u_{i_2}, & q_i^0 = - \ln(u_{i_3} u_{i_4}), \nonumber \\
q_i^x = q_i^0 \sqrt{1- c_i^2} \cos \varphi_i, & q_i^y = q_i^0 \sqrt{1-c_i^2} \sin \varphi_i, & q_i^z = q_i^0 c_i.
\end{eqnarray}
\item The four-vectors $q_i^\mu$ are then transformed into the four-vectors $p_i^\mu$, using the transformation
(\ref{boost_and_scale}).
\end{enumerate}
Each event has the uniform weight
\bq
\label{masslessweight}
w_0 & = & (2 \pi)^{4-3n} \left( \frac{\pi}{2} \right)^{n-1} \frac{(P^2)^{n-2}}{\Gamma(n) \Gamma(n-1)}.
\eq
Phase-space configurations corresponding to massive particles can be generated
by starting from a massless configuration and then transforming this configuration into
one with the desired masses. This is done as follows : Let $p_i^\mu$ be a set of massless momenta.
One starts again from the phase-space integral for massless particles
\begin{eqnarray}
\Phi_n(\{p\}) & = & \int \prod\limits_{i=1}^n \frac{d^4 p_i}{(2 \pi)^3} \theta(p_i^0) \delta(p_i^2)  
(2 \pi)^4 \delta^4\left( P - \sum\limits_{i=1}^n p_i \right).
\end{eqnarray}
The $p_i^\mu$ are transformed into the four-momenta $k_i^\mu$ as follows :
\begin{eqnarray}
\label{massive}
k_i^0 = \sqrt{ m_i^2 + \xi^2 (p_i^0)^2 }, & & 
\vec{k}_i = \xi \vec{p}_i,
\end{eqnarray}
where $\xi$ is a solution of the equation
\begin{eqnarray}
\label{eta}
\sqrt{P^2} & = & \sum\limits_{i=1}^n \sqrt{ m_i^2 + \xi^2 (p_i^0)^2 }.
\end{eqnarray}
It should be noted that in the general case no analytic expression for $\xi$ exists
and $\xi$ has to be computed numerically.
After some manipulations one arrives at
\begin{eqnarray}
\Phi_n(\{p\}) & = & \int \prod\limits_{i=1}^n \frac{d^4 k_i}{(2 \pi)^3} \theta(k_i^0) \delta(k_i^2-m^2_i)  
(2 \pi)^4 \delta^4\left( P - \sum\limits_{i=1}^n k_i \right) \cdot W(\{p\},\{k\}), \nonumber \\
\end{eqnarray}
where the weight is given by
\begin{eqnarray}
\label{massiveweight}
w_m & = & (P^2)^{2-n} \left( \sum\limits_{i=1}^n |\vec{k}_i| \right)^{2n-3}
\left( \prod\limits_{i=1}^n \frac{|\vec{k}_i|}{k_i^0} \right)
\left( \sum\limits_{i=1}^n \frac{|\vec{k}_i|^2}{k_i^0} \right)^{-1}.
\end{eqnarray}
In contrast to the massless case, the weight is no longer constant but varies over phase space.\\
\\
To generate events with massive particles in the final state one therefore proceeds as follows:
\begin{enumerate}
\item Generate an event with $n$ massless particles.
\item Solve eq.~\ref{eta} numerically.
\item Use eq.~\ref{massive} to obtain the momenta of the massive particles.
\end{enumerate}
The weight of such an event is then given by
\bq
w & = & w_m w_0
\eq
with $w_0$ and $w_m$ defined in eq.~\ref{masslessweight} and eq.~\ref{massiveweight}, respectively.

\subsubsection{Generating configurations close to soft or collinear regions}

In massless QCD individual terms in the leading order squared matrix element are of the form
\bq
{\cal M} & = & \frac{m}{s_{i_1 i_2} s_{i_2 i_3} ... s_{i_{k-1} i_k}}.
\eq
Integration over phase space leads to collinear singularities when one of the $s_{ij}$ approaches
zero, and to soft singularities when two adjacent invariants $s_{i_{j-1} i_j}$ and  
$s_{i_j} s_{i_{j+1}}$ 
approach zero simultaneously. Event generators use therefore a cutoff $s_{min}$ to avoid this region.
In perturbative calculations these singularities cancel against similar divergences encountered
in loop integrals, but the cancelation occurs only after the phase space integration
has been performed. On the other hand the phase space integrations have to be done numerically.
Two approaches exist to handle this situation: phase space slicing and the subtraction method.
Within the slicing approach the phase space is split into a region $s>s_{min}$, where numerical
integration is performed, and a small region $s<s_{min}$, where after approximation of the integrand
one integration can be performed analytically, and the resulting singularities cancel then against
the corresponding ones in the loop amplitudes.
Within the subtraction method one adds and subtracts a suitable chosen term, such that each integration
(e.g. the integration over the real emission part and the integration over the virtual correction) 
is finite.
Nevertheless it will also be
useful within the subtraction method  
to split the integration in the real emission part in two pieces in order to improve the convergence of the Monte Carlo integration:
One region $s<s_{min}$ which will give a small or negligible contribution and a region
$s>s_{min}$.\\
\\
In short in any case we are dealing with integrals of the form $\int ds f(s)/s$, with lower boundary
$s_{min}$. To improve the efficiency of the Monte Carlo integration it is desirable to remap
the $1/s$-behaviour of the integrand. In the simple example above this could be done
by a change of variables $y=\ln s$. In real problems there is more than one invariant $s$
and the remapping is done by relating a $(n+1)$-parton configuration with one
soft or collinear parton such that $s_{as} s_{sb}$ is the smallest product of all adjacent
products
to a ``hard'' $n$-parton configuration \cite{mercutio}. 
In the region where $s_{as} s_{sb}$ is the smallest product, we remap the phase space as follows:
Let $k_a'$, $k_s$ and $k_b'$ be the corresponding momenta such that $s_{as} = (k_a' + k_s)^2$, 
$s_{sb} = (k_b' + k_s)^2$ and $s_{ab} = (k_a' + k_s + k_b')^2$. We want to relate this $(n+1)$ particle
configuration to a nearby ``hard'' $n$-particle configuration with $(k_a + k_b)^2 = (k_a' + k_s + k_b')^2$,
where $k_a$ and $k_b$ are the corresponding ``hard'' momenta.
Using the factorization of the phase space, we have 
\begin{eqnarray}
d\Phi_{n+1} & = & d\Phi_{n-1} \frac{dK^2}{2 \pi} d\Phi_3(K,k_a',k_s,k_b').
\end{eqnarray}
The three-particle phase space is given
by
\begin{eqnarray}
d\Phi_3(K,k_a',k_s,k_b') & = & \frac{1}{32 (2 \pi)^5 s_{ab}} ds_{as} ds_{sb} d\Omega_b' d\phi_s \nonumber \\
& = & \frac{1}{4 (2 \pi)^3 s_{ab}} ds_{as} ds_{sb} d\phi_s d\Phi_2(K,k_a,k_b)
\end{eqnarray}
and therefore
\begin{eqnarray}
d\Phi_{n+1} & = & d\Phi_n \frac{ds_{as} ds_{sb} d\phi_s}{4 (2 \pi)^3 s_{ab} }.
\end{eqnarray}
The region of integration for $s_{as}$ and $s_{sb}$ is 
$s_{as} > s_{min}$, $s_{sb} > s_{min}$ (since we want to restrict the integration to the region
where the invariants are larger than $s_{min}$)
and $s_{as} + s_{sb} < s_{ab}$ (Dalitz plot for massless particles).
It is desirable to absorb poles in $s_{as}$ and $s_{sb}$ into the measure. 
A naive numerical integration of these poles without any remapping
results in a poor accuracy.
This is done
by changing the variables according to
\begin{eqnarray}
s_{as} = s_{ab} \left( \frac{s_{min}}{s_{ab}} \right)^{u_1}, & & 
s_{sb} = s_{ab} \left( \frac{s_{min}}{s_{ab}} \right)^{u_2},
\end{eqnarray}
where $0 \leq u_1, u_2 \leq 1$. Note that $u_1, u_2 > 0$ enforces $s_{as}, s_{sb} > s_{min}$.
Therefore this transformation of variables may only be applied to invariants $s_{ij}$ 
where the region $0 < s_{ij} < s_{min}$ is cut out.
The phase space measure becomes
\begin{eqnarray}
d\Phi_{n+1} & = & d\Phi_{n} \frac{1}{4 (2 \pi)^3} \frac{s_{as} s_{sb}}{s_{ab}} \ln^2\left(\frac{s_{min}}{s_{ab}}\right)
\Theta(s_{as} + s_{sb} < s_{ab} ) du_1 du_2 d\phi_s .
\end{eqnarray}
This give the following algorithm for generating a $(n+1)$-parton configuration:
\begin{enumerate}
\item Take a ``hard'' $n$-parton configuration and pick out two momenta $k_a$ and $k_b$. Use three
uniformly distributed random number $u_1,u_2,u_3$ and set
\begin{eqnarray}
s_{ab} & = & (k_a + k_b)^2 ,\nonumber \\
s_{as} & = & s_{ab} \left( \frac{s_{min}}{s_{ab}} \right)^{u_1} ,\nonumber \\
s_{sb} & = & s_{ab} \left( \frac{s_{min}}{s_{ab}} \right)^{u_2} ,\nonumber \\
\phi_s & = & 2 \pi u_3.
\end{eqnarray}
\item If $(s_{as} + s_{sb} ) > s_{ab} $, reject the event.
\item If not, solve for $k_a'$, $k_b'$ and $k_s$.
If $s_{as} < s_{sb}$ we want to have $k_b' \rightarrow k_b$ as $s_{as} \rightarrow 0$.
Define
\begin{eqnarray}
E_a = \frac{s_{ab} - s_{sb}}{2 \sqrt{s_{ab}}}, \;\;\;
E_s = \frac{s_{as}+s_{sb}}{2 \sqrt{s_{ab}}} ,\;\;\;
E_b = \frac{s_{ab} - s_{as}}{2 \sqrt{s_{ab}}}, 
\end{eqnarray}
\begin{eqnarray}
\theta_{ab}  =  \arccos \left( 1 -\frac{s_{ab} - s_{as} - s_{sb}}{2 E_a E_b} \right), & &
\theta_{sb}  =  \arccos \left( 1 - \frac{s_{sb}}{2 E_s E_b} \right) .
\end{eqnarray}
It is convenient to work in a coordinate system which is obtained by a Lorentz transformation to the
center of mass of $k_a+k_b$ and a rotation such that $k_b'$ is along the positive $z$-axis. In that
coordinate system
\begin{eqnarray}
p_a' & = & E_a ( 1, \sin \theta_{ab} \cos(\phi_s+\pi), \sin \theta_{ab} \sin(\phi_s+\pi), \cos \theta_{ab} ) ,\nonumber \\
p_s & = & E_s ( 1, \sin \theta_{sb} \cos \phi_s, \sin \theta_{sb} \sin \phi_s, \cos \theta_{sb} ) ,\nonumber \\
p_b' & = & E_b ( 1, 0, 0, 1) .
\end{eqnarray}
The momenta $p_a'$, $p_s$ and $p_b'$ are related to the momenta $k_a'$, $k_s$ and $k_b'$ by a sequence of
Lorentz transformations back to the original frame
\begin{eqnarray}
k_a' & = & \Lambda_{boost} \Lambda_{xy}(\phi) \Lambda_{xz}(\theta) p_a'
\end{eqnarray}
and analogous for the other two momenta. 
The explicit formulae for the Lorentz transformations are obtained as follows :\\
\\
Denote by $K = \sqrt{(k_a+k_b)^2}$ and by $p_b$ the coordinates of the hard momentum $k_b$ in the center of
mass system of $k_a+k_b$. $p_b$ is given by
\begin{eqnarray}
p_b & = & \left( 
\frac{Q^0}{K} k_b^0 - \frac{\vec{k}_b \cdot \vec{Q}}{K}, \vec{k}_b + \left( \frac{\vec{k}_b \cdot \vec{Q}}{K (Q^0+K)} 
  - \frac{k_b^0}{K} \right) \vec{Q}
\right)
\end{eqnarray}
with $Q = k_a + k_b$. The angles are then given by
\begin{eqnarray}
\theta & = & \arccos \left( 1 - \frac{p_b \cdot p_b'}{2 p_b^t p_b^{t'}} \right), \nonumber \\
\phi & = & \arctan\left( \frac{p_b^y}{p_b^x} \right).
\end{eqnarray}
The explicit form of the rotations is
\begin{eqnarray}
\Lambda_{xz}(\theta) & = & \left(
\begin{array}{cccc}
1 & 0 & 0 & 0 \\
0 & \cos \theta & 0 & \sin \theta \\
0 & 0 & 1 & 0 \\
0 & - \sin \theta & 0 & \cos \theta \\
\end{array}
\right), \nonumber \\
\Lambda_{xy} (\phi) & = & 
\left(
\begin{array}{cccc}
1 & 0 & 0 & 0 \\
0 & \cos \phi & - \sin \phi & 0 \\
0 & \sin \phi & \cos \phi & 0 \\
0 & 0 & 0 & 1 \\
\end{array}
\right).
\end{eqnarray}
The boost $k' = \Lambda_{boost} q $ is given by 
\begin{eqnarray}
k' & = & \left( 
\frac{Q^0}{K} q^0 + \frac{\vec{q} \cdot \vec{Q}}{K}, \vec{q}+ \left( \frac{\vec{q} \cdot \vec{Q}}{K (Q^0+K)} 
  + \frac{q^0}{K} \right) \vec{Q}
\right)
\end{eqnarray}
with $Q = k_a + k_b$ and $K = \sqrt{(k_a+k_b)^2}$.
\item If $s_{as} > s_{sb}$, exchange $a$ and $b$ in the formulae above.
\item The ``soft'' event has then the weight
\begin{eqnarray}
w_{n+1} & = & \frac{\pi}{2} \frac{1}{(2 \pi)^3} \frac{s_{as} s_{sb}}{s_{ab}} \ln^2 \left( \frac{s_{min}}{s_{ab}} \right) w_n,
\end{eqnarray}
where $w_n$ is the weight of the original ``hard'' event.
\end{enumerate}

\subsection{Summary on generating specific samples}

In this section we presented some methods to generate samples according to a specified
distribution. If the inverse of the cumulative distribution function is known, the inverse
transform method can be used, otherwise one often relies on the acceptance-rejection method.
For high-dimensional problems the Metropolis algorithm is a popular choice. It is often
applied to spin glasses and to lattice field theories.\\
\\
In perturbative calculations or event generators for high energy physics
one often wants to sample the phase space of the outgoing particles. We reviewed the standard
sequential approach, the democratice approach and a method for generating soft and collinear
particles.
The sequential approach has an advantage if peaks due to massive particles occur
in intermediate stages of the process. By choosing the $M_i^2$ cleverly, VEGAS can take
the invariant mass of the resonance along one axis and easily adapt to the peak.
The democratic approach has the advantage that each event has a uniform weight and has
its application in massless theories.
The method for generating soft and collinear particles is, as the name already indicates,
designed to generate efficiently phase space configurations, where one particle
is soft or collinear. This is important for example in NLO calculations
where otherwise most computer time is spent in calculating the contributions from the 
real emission part.\\ 
\\
Further reading: The lecture notes by W. Krauth \cite{krauth} and A.D. Sokal \cite{sokal} deal
with applications of Monte Carlo methods in statistical physics and are worth reading.
Further there are review articles on the Metropolis algorithm by G. Bhanot \cite{bhanot} and
on the Potts model by F.Y. Wu \cite{potts}.
There are many introductions to lattice field theory, I borrowed the ideas from the review
article by R.D. Kenway \cite{kenway}. The book by Byckling and Kajantie \cite{kajantie} is the standard book
on phase space for final state particles.\\
\\
\bs
{\it Exercise \theexercise: Application of the Potts model to data clustering. This technique
is due to M. Blatt, S. Wiseman and E. Domany \cite{flowers}.
Given a set of data (for example as points in a $d$-dimensional vector space), 
cluster analysis tries to group this data into clusters, such that the members
in each cluster are more similar to each other than to members of different
clusters. An example is an measurement of four different quantities from 150 Iris flowers.
(In this specific example it was known that each flower belonged to exactly one subspecies
of the Iris: either Iris Setosa, Iris Versicolor or Iris Virginica.
The four quantities which were measured are the petal length, the petal width, the sepal length
and the sepal width.)
Given the data points alone, the task is to decide which flowers belong to the same subspecies.)
M. Blatt et al. suggested to use the Potts model for this problem as follows:
Given $N$ points $x_i$ in a $d$-dimensional vector space one chooses a value $q$ for the number
of different spin states and a model for the spin-spin couplings $J_{ij}=f(d_{ij})$, where
$d_{ij}$ is the distance between the points $x_i$ and $x_j$ and $f$ a function depending on the 
distance $d_{ij}$. To minimize computer time one usually
chooses $f(d_{ij})=0$ for $d_{ij}>d_{cut}$.
One then performs a Monte Carlo simulation of the corresponding Potts model. At low temperature
all spins are aligned, where as at high temperature the orientation of the spins is random.
In between there is a super-paramagnetic phase, where spins are aligned inside ``grains'', but the
orientation of different grains is random.
Using the Swendsen-Wang algorithm one first tries to find this super-paramagnetic phase by
monitoring the magnetic susceptibility (there should be a strong peak in $\chi$ at the
ferromagnetic-superparamagnetic phase transition, furthermore, at higher temperature there will
be a significant drop in the susceptibility where the alignment of the spins inside a grain
break up.)
Having found the right temperature one then calculates the probability that two neighbouring sides
are in one Swendsen-Wang cluster.
If this probability is greater than $1/2$ they are assigned to the same data cluster.
The final result should depend only mildly on the 
exact value of $q$ and the functional form of $f(d_{ij})$. This can be verified by performing the simulation
with different choices for these parameters.
Write a Monte Carlo simulation which clusters the data according to the algorithm outlined above.
You can find the data for the Iris flowers on the web, for example
at http://www.math.uah.edu/stat/data/Fisher.html.
\stepcounter{exercise}}
\es

\begin{appendix}

\section{Orthogonal polynomials}
\label{orthogonal}
We list here some properties of the most common known orthogonal polynomials
(Legendre, Tschebyscheff, Gegenbauer, Jacobi, Laguerre and Hermite).
For each set we state the standard normalization and the differential equation
to which the polynomials are solutions. We further give the explicit expression,
the recurrence relation as well as the generating function and Rodrigues' formula.
The information is taken from the book by A. Erd\'elyi \cite{erdelyi}.

\subsection{Legendre polynomials}

The Legendre polynomials $P_n(x)$ are defined on the interval $[-1,1]$ with weight function
$w(x)=1$. The standard normalization is
\bq
\int\limits_{-1}^1 dx P_n(x) P_m(x) & = & \frac{2}{2n+1} \delta_{nm}.
\eq
The Legendre polynomials are solutions of the differential equation
\bq
(1-x^2) y'' -2 x y' + n (n+1) y & = & 0.
\eq
The explicit expression is given by
\bq
P_n(x) & = & \frac{1}{2^n} \sum\limits_{m=0}^{[n/2]} (-1)^m
\left( \begin{array}{c} n \\ m \\ \end{array} \right)
\left( \begin{array}{c} 2n-2m \\ n \\ \end{array} \right)
x^{n-2m},
\eq
where $[n/2]$ denotes the largest integer smaller or equal to $n/2$.
They can also be obtained through the
recurrence relation:
\bq
& & P_0(x) = 1, \;\;\;
P_1(x) = x, \nonumber \\
& & (n+1) P_{n+1}(x) = (2 n + 1) x P_n(x) - n P_{n-1}(x).
\eq
Alternatively one may use the generating function
\bq
\frac{1}{\sqrt{1 -2 x z + z^2}} & = & \sum\limits_{n=0}^\infty P_n(x) z^n, \;\;\; -1 \leq x \leq 1,
\;\; |z| < 1,
\eq
or Rodrigues' formula :
\bq
P_n(x) & = & \frac{1}{2^n n!} \frac{d^n}{dx^n} (x^2-1)^n.
\eq

\subsection{Tschebyscheff polynomials}

The Tschebyscheff polynomials of the first kind $T_n(x)$ are defined on the interval $[-1,1]$ with the
weight function
$w(x)=1/\sqrt{1-x^2}$.
They can be viewed as a special case of the Gegenbauer polynomials. Due to their special
normalization we discuss them separately.
They are normalized as
\bq
\int\limits_{-1}^1 dx (1-x^2)^{\mu-1/2} T_n(x) T_m(x) & = &
\left\{ \begin{array}{cc} \pi/2, & n \neq 0, \\ \pi, & n = 0.  \\ \end{array} \right.
\eq
The Tschebyscheff polynomials are solutions of the differential equation
\bq
(1-x^2) y'' - x y' + n^2 y & = & 0.
\eq
The explicit expression reads
\bq
T_n(x) & = & \frac{n}{2} \sum\limits_{m=0}^{[n/2]}
(-1)^m \frac{(n-m-1)!}{m!(n-2m)!} (2x)^{n-2m}.
\eq
The recurrence relation is given by
\bq
& & T_0(x) = 1, \;\;\;
T_1(x) = x, \nonumber \\
& & T_{n+1}(x) = 2 x T_n(x) - T_{n-1}(x).
\eq
The generating function is
\bq
\frac{1-xz}{1 -2 x z + z^2} & = & \sum\limits_{n=0}^\infty T_n(x) z^n, \;\;\; -1 \leq x \leq 1,
\;\; |z| < 1.
\eq
Rodrigues' formula reads
\bq
T_n(x) & = & \frac{(-1)^n (1-x^2)^{1/2} \sqrt{\pi} }
{2^{n+1} \Gamma\left(n+\frac{1}{2}\right)}
\frac{d^n}{dx^n} \left(1-x^2\right)^{n-1/2}.
\eq

\subsection{Gegenbauer polynomials}

The Gegenbauer polynomials $C^\mu_n(x)$ are defined on the interval $[-1,1]$ with the
weight function
$w(x) = (1-x^2)^{\mu-1/2}$ for $\mu > -1/2$.
The standard normalization is
\bq
\label{gegenbauernormalization}
\int\limits_{-1}^1 dx (1-x^2)^{\mu-1/2} C_n^\mu(x) C_m^\mu(x) & = &
\frac{\pi 2^{1-2\mu} \Gamma(n+2\mu)}{n! (n+\mu) \left(\Gamma(\mu)\right)^2} \delta_{nm}.
\eq
The Gegenbauer polynomials are solutions of the differential equation
\bq
(1-x^2) y'' - ( 2 \mu + 1 ) x y' + n (n+2\mu) y & = & 0.
\eq
The explicit expression reads
\bq
C_n^\mu(x) & = & \frac{1}{\Gamma(\mu)} \sum\limits_{m=0}^{[n/2]}
(-1)^m \frac{\Gamma(\mu+n-m)}{m! (n-2m)!}(2x)^{n-2m}, 
\eq
The recurrence relation is given by
\bq
& & C_0^\mu(x) = 1, \;\;\;
C_1^\mu(x) = 2 \mu x, \nonumber \\
& & (n+1) C_{n+1}^\mu(x) = 2 ( n+\mu)x C_n^\mu(x) - (n + 2 \mu -1) C_{n-1}^\mu(x).
\eq
The generating function is
\bq
\frac{1}{\left(1 -2 x z + z^2\right)^\mu} & = & \sum\limits_{n=0}^\infty C_n^\mu(x) z^n, \;\;\; -1 \leq x \leq 1,
\;\; |z| < 1.
\eq
Rodrigues' formula reads
\bq
C_n^\mu(x) & = & \frac{(-1)^n 2^n n! \Gamma(2\mu) \Gamma\left(\mu+n+\frac{1}{2}\right)}
{\Gamma\left(\mu+\frac{1}{2}\right) \Gamma\left(n+2\mu\right)(1-x^2)^{\mu-1/2}}
\frac{d^n}{dx^n} \left(1-x^2\right)^{n+\mu-1/2}.
\eq
Special cases : For $\mu = 1/2$ the Gegenbauer polynomials reduce to the Legendre polynomials, e.g
$C_n^{1/2}(x)=P_n(x)$.\\
\\
The polynomials $C^1_n(x)$ are called Tschebyscheff polynomials of the second kind and denoted
by $U_n(x)$.\\
\\
The case $\mu=0$ corresponds to Tschebyscheff polynomials of the first kind.
They cannot be normalized accoring to eq.~\ref{gegenbauernormalization}
and have been treated separately above.

\subsection{Jacobi polynomials}

The Jacobi polynomials $P_n^{(\alpha,\beta)}(x)$ are defined on the interval $[-1,1]$ with
the weight function
$w(x) = (1-x)^\alpha (1+x)^\beta$
for $\alpha,\beta > -1$.
The standard normalization is given by 
\bq
\int\limits_{-1}^1 dx (1-x)^\alpha ( 1+x)^\beta P_n^{(\alpha,\beta)}(x) P_m^{(\alpha,\beta)}(x)
  & = & \frac{2^{\alpha+\beta+1} \Gamma(n+\alpha+1)
\Gamma(n+\beta+1)}{(2n+\alpha+\beta+1) n! \Gamma(n+\alpha+\beta+1)} \delta_{nm}. \nonumber \\
\eq
Differential equation:
\bq
(1-x^2) y'' + ( \beta-\alpha-(\alpha+\beta+2) x ) y' + n
(n+\alpha+\beta+1) y & =& 0
\eq
Explicit expression:
\bq
P_n^{(\alpha,\beta)}(x) & = & \frac{1}{2^n} \sum\limits_{m=0}^n
\left( \begin{array}{c} n+\alpha \\ m \\ \end{array} \right)
\left( \begin{array}{c} n+\beta \\ n-m \\ \end{array} \right)
(x-1)^{n-m} (x+1)^m
\eq
Recurrence relation :
\bq
P^{(\alpha,\beta)}_0(x) = 1, \;\;\;
P^{(\alpha,\beta)}_1(x) = \left( 1 + \frac{1}{2} \left( \alpha + \beta \right) \right) x + \frac{1}{2} \left( \alpha - \beta \right), \nonumber 
\eq
\bq
\lefteqn{
2 \left( n + 1 \right) \left( n + \alpha + \beta + 1 \right) \left( 2 n + \alpha + \beta \right) P^{(\alpha,\beta)}_{n+1}(x) =} &  & \nonumber \\
& = & \left( 2 n + \alpha +\beta + 1 \right) \left( \left( \alpha^2 - \beta^2 \right) + \left( 2 n + \alpha + \beta + 2 \right) \left( 2 n + \alpha +\beta \right) x \right) P^{(\alpha,\beta)}_n(x) \nonumber \\
& & - 2 \left( n+ \alpha \right) \left( n + \beta \right) \left( 2 n + \alpha + \beta + 2 \right) P^{(\alpha,\beta)}_{n-1}(x).
\eq
Generating function:
\bq
\lefteqn{
R^{-1} (1-z+R)^{-\alpha} (1+z+R)^{-\beta} = \sum\limits_{n=0}^\infty
2^{-\alpha-\beta}P_n^{(\alpha,\beta)}(x) z^n,} & & \nonumber  \\
& & R = \sqrt{1-2xz+z^2}, \;\;\; -1 \leq x \leq 1, \;\;\; |z| < 1.
\eq
Rodrigues' formula:
\bq
P_n^{(\alpha,\beta)}(x) &= & \frac{(-1)^n}{2^n n! (1-x)^\alpha (1+x)^\beta}
\frac{d^n}{dx^n} \left( (1-x)^{n+\alpha} (1+x)^{n+\beta} \right)
\eq
In the case $\alpha=\beta$ the Jacobi polynomials reduce to the Gegenbauer
polynomials with $\mu=\alpha+1/2$. The exact relation is
\bq
P_n^{(\alpha,\alpha)}(x) & = & 
\frac{\Gamma(2\alpha+1)}{\Gamma(2\alpha+1+n)} 
\frac{\Gamma(\alpha+1+n)}{\Gamma(\alpha+1)} C_n^{\alpha+\frac{1}{2}}(x).
\eq

\subsection{Generalized Laguerre polynomials}

The generalized Laguerre polynomials $L_n^{(\alpha)}(x)$ are defined on the interval
$[0,\infty]$ with the weight function 
$w(x) = x^\alpha e^{-x}$ for $a > -1$.
The standard normalization is given by
\bq
\int\limits_0^\infty dx x^\alpha e^{-x} L_n^{(\alpha)}(x) L_m^{(\alpha)}(x)
 & = & \frac{\Gamma(n+\alpha+1)}{n!} \delta_{nm}.
\eq
Differential equation:
\bq
x y'' + (\alpha + 1 -x) y' + n y & = & 0
\eq
Explicit expression:
\bq
L_n^{(\alpha)} & = & \sum\limits_{m=0}^n
\frac{(-1)^m}{m!}
\left( \begin{array}{c} n+ \alpha \\ n-m \\ \end{array} \right)
x^m
\eq
Recurrence relation:
\bq
& & L_0^{(\alpha)}(x) = 1, \;\;\;
L_1^{(\alpha)}(x) = 1 + \alpha -x, \nonumber \\
& & (n+1) L_{n+1}^{(\alpha)}(x) = \left( ( 2 n + \alpha +1) - x \right)
L_n^{(\alpha)}(x) - (n+\alpha)L_{n-1}^{(\alpha)}(x).
\eq
Generating function:
\bq
(1-z)^{-\alpha-1} \exp \left( \frac{xz}{z-1} \right)
& =& \sum\limits_{n=0}^\infty L_n^{(\alpha)}(x) z^n, \;\;\; |z| < 1.
\eq
Rodrigues' formula:
\bq
L_n^{(\alpha)}(x) & = & \frac{1}{n! x^\alpha e^{-x}}
\frac{d^n}{dx^n} \left( x^{n+\alpha} e^{-x} \right)
\eq
Special cases : The polynomials $L^{(0)}_n(x)$ are the Laguerre polynomials and denoted
by $L_n(x)$.

\subsection{Hermite polynomials}

The Hermite polynomials $H_n(x)$ are defined on the interval $[-\infty,\infty]$ with the weight
function
$w(x) = e^{-x^2}$.
The Hermite polynomials appear in the quantum mechanical harmonic oscillator.
The standard normalization is given by
\bq
\int\limits_{-\infty}^\infty dx e^{-x^2} H_n(x) H_m(x) & = & 2^n \sqrt{\pi}
n! \delta_{n m}.
\eq
Differential equation:
\bq
y'' - 2 x y' + 2 n y & = & 0
\eq
The explicit expression is given by
\bq
H_n(x) & = & n! \sum\limits_{m=0}^{[n/2]} (-1)^m \frac{(2x)^{n-2m}}{m!(n-2m)!}.
\eq
Recurrence relation:
\bq
& & H_0(x) = 1, \;\;\;
H_1(x) = 2 x, \nonumber \\
& & H_{n+1}(x) = 2 x H_n(x) -2 n H_{n-1}(x).
\eq
Generating function:
\bq
e^{-t^2+2 x t} & = & \sum\limits_{n=0}^\infty \frac{1}{n!} H_n(x) t^n
\eq
Rodrigues' formula:
\bq
H_n(x) & = & (-1)^n e^{x^2} \frac{d^n}{dx^n} e^{-x^2}
\eq

\section{Sampling some specific distriubtions}

We list here some cooking recipes how to generate samples according to some
special distributions (Gaussian, $\chi^2$, binomial, Poisson, gamma, beta and Student t).
The algorithms are taken from Ahrens and Dieter \cite{distribution} and the particle data group \cite{pdg}.

\subsection{Gaussian distribution}

A random variable distributed according to the probability density function
\bq
p(x) & = & \frac{1}{\sqrt{2\pi \sigma^2}} e^{-\frac{(x-\mu)^2}{2 \sigma^2}}
\eq
is called normal or Gaussian distributed. As already mentioned in 
sect.~\ref{application}
the Box-Muller algorithm allows one to generate two independent
variables $x_1$ and $x_2$, distributed according to a Gaussian 
distribution with mean $\mu=0$ and variation $\sigma^2=1$ from two
independent variables $u_1$ and $u_2$, uniformly distributed in
$[0,1]$ as follows:
\bq
x_1 & = & \sqrt{-2 \ln u_1} \cos(2 \pi u_2), \nonumber \\
x_2 & = & \sqrt{-2 \ln u_1} \sin(2 \pi u_2). 
\eq

\subsection{$\chi^2$-distribution}

If $x_1$, ..., $x_n$ are $n$ independent Gaussian random variables
with mean $\mu_i$ and variance $\sigma_i^2$ for the variable $x_i$,
the sum
\bq
x & = & \sum\limits_{i=1}^n \frac{(x_i-\mu_i)^2}{\sigma_i^2}
\eq
is distributed as a $\chi^2$ with $n$ degrees of freedom.
The $\chi^2(n)$ distribution has the probability density function
\bq
p(x) & = & \frac{x^{\frac{n}{2}-1} e^{-\frac{x}{2}}}
{2^{\frac{n}{2}} \Gamma \left( \frac{n}{2} \right)},
\eq
where $x \ge 0$ is required.  
The $\chi^2(n)$-distribution has mean $n$ and variance $2n$.
To generate a random variable $x$ which is distributed as $\chi^2(n)$
one can start from the definition and generate $n$ Gaussian random
variables.
However there is a more efficient approach. For $n$ even one
generates $n/2$ uniform numbers $u_i$ and sets
\bq
x & = & -2 \ln \left( u_1 u_2 \cdot ... \cdot u_{n/2} \right).
\eq
For $n$ odd, one generates $(n-1)/2$ uniform numbers $u_i$ and
one Gaussian $y$ and sets
\bq
x & = & -2 \ln \left( u_1 u_2 \cdot ... \cdot u_{(n-1)/2} \right) + y^2.
\eq

\subsection{Binomial distribution}

Binomial distributions are obtained from random processes with exactly
two possible outcomes. If the propability of a hit in each trial is
$p$, then the probability of obtaining exactly $r$ hits in $n$ trials
is given by
\bq
\label{binomial}
p(r) & = & \frac{n!}{r! (n-r)!} p^r (1-p)^{n-r},
\eq
where $r=0,1,2,3,...$ is an integer and $0 \le p \le 1$.
A random variable $r$ distributed according to eq.~\ref{binomial}
is called binomialy distributed. The binomial distribution
has mean $np$ and variance $np(1-p)$.
One possibility to generate integer numbers $r=0,...,n$, which are distributed according to a binomial
distribution, is directly obtained from the definition:
\begin{enumerate}
\item Set $r=0$ and $m=0$.
\item Generate a uniform random number $u$. If $u \le p$ increase $k=k+1$.
\item Increase $m=m+1$. If $m < n$ go back to step 2, otherwise return $k$.
\end{enumerate}
The computer time for this algorithm grows linearly with $n$.
The algorithm can be used for small values of $n$. For large $n$ there are better algorithms 
\cite{distribution,pdg}. 

\subsection{Poisson distirubtion}

The Poisson distribution is the limit $n \rightarrow \infty$,
$p \rightarrow 0$, $n p = \mu$ of the binomial distribution.
The probability density function for the Poisson distribution
reads
\bq
p(r) & = & \frac{\mu^r e^{-\mu}}{r!},
\eq
where $\mu>0$ and $r=0,1,2,3,...$ is an integer.
The Poisson distribution has mean $\mu$ and variance $\mu$.
For large $\mu$ the Poisson distribution approaches the Gaussian distribution.
To generate integer numbers $r=0,1,2,...$  distributed according to a Poisson distribution,
one proceeds as follows:
\begin{enumerate}
\item Initialize $r=0$, $A=1$.
\item Generate a uniform random number $u$, set $A=uA$. If $A \le e^{-\mu}$ accept $r$ and stop.
\item Increase $r=r+1$ and goto step 2.
\end{enumerate}

\subsection{Gamma distribution}

The probability density function of the gamma distribution is given by
\bq
p(x) & = & \frac{x^{k-1} \lambda^k e^{-\lambda x}}{\Gamma(k)}
\eq
with $x > 0$, $\lambda > 0$ and $k>0$ is not required to be an integer.
The gamma distribution has mean $k/\lambda$ and variance
$k/\lambda^2$.\\
\\
The special case $k=1$ gives the exponential distribution
\bq
p(x) & = & \lambda e^{-\lambda x}.
\eq
To generate random numbers distributed according to a gamma distribution, we consider
the cases $k=1$, $k>1$ and $0<k<1$ separately. For simplicity we use $\lambda=1$. Results
for $\lambda \neq 1$ are easily obtained by dividing the resulting random number $x$
by $\lambda$.\\
\\
The case $k=1$ is just the exponential distribution. One generates a uniform random number
$u$ and sets $x=-\ln(u)$.\\
\\
For the case $0<k<1$ one uses a rejection technique: One first observes
that the function
\bq
g(x) & = & \left\{ \begin{array}{ll}
\frac{x^{k-1}}{\Gamma(k)}, & 0 \le x \le 1, \\
\frac{e^{-x}}{\Gamma(k)}, & 1 \le x. \\
\end{array}
\right.
\eq
is a majorizing function for $p(x)$. The function
\bq
h(x) & = & \left\{ \begin{array}{ll}
\frac{e k}{e+k} x^{k-1}, & 0 \le x \le 1, \\
\frac{e k}{e+k} e^{-x}, & 1 \le x. \\
\end{array}
\right.
\eq
with $e=2.71828...$ being the base of the natural logarithm
is a probability density proportional to $g(x)$, which can easily be sampled.
This yields the following algorithm:
\begin{enumerate}
\item Set $v_1=1+k/e$.
\item Generate two uniform random number $u_1$, $u_2$ and set $v_2=v_1 u_1$.
\item If $v_2 \le 1$, set $x=v_2^{1/k}$ and accept $x$ if $u_2 \le e^{-x}$,
otherwise go back to step 2.
\item If $v_2 > 1$, set $x=-\ln((v_1-v_2)/k)$ and accept $x$ if $u_2\le x^{k-1}$,
otherwise go back to step 2.
\end{enumerate}
For the case $k>1$ one uses the majorization
\bq
\frac{x^{k-1} e^{-x}}{\Gamma(k)} & \le & \frac{1}{\Gamma(k)} \frac{(k-1)^{k-1} e^{-(k-1)}}{1
+\frac{\left(x-(k-1) \right)^2}{2k-1}}, \;\;\;\; k > 1,
\eq
and obtains the following algorithm:
\begin{enumerate}
\item Set $b=k-1$, $A=k+b$ and $s=\sqrt{A}$.
\item Generate a uniform random number $u_1$ and set
$t=s \tan \left( \pi(u_1-1/2) \right)$ and $x=b+t$.
\item If $x<0$ go back to step 2.
\item Generate a uniform random number $u_2$ and accept $x$ if
\bq
u_2 & \le & \exp \left( b \ln \left(\frac{x}{b} \right) -t +\ln \left( 1 + \frac{t^2}{A} \right) \right),
\eq 
otherwise go back to step 2.
\end{enumerate}

\subsection{Beta distributions}

The probability density function of the beta distribution is given
by
\bq
p(x) & = & \frac{x^{\alpha-1} (1-x)^{\beta-1}}{B(\alpha,\beta)}, \;\;\;\; 0\le x \le 1, \;\;\; \alpha,\beta >0,
\eq
where $B(\alpha,\beta)=\Gamma(\alpha) \Gamma(\beta) / \Gamma(\alpha+\beta)$.
Samples according to a beta distribution can be obtained from a gamma distribution.
If $x$ and $y$ are two independent deviates which are (standard, e.g. $\lambda=1$) gamma-distributed
with parameters $\alpha$ and $\beta$ respectively, then $x/(x+y)$ is $B(\alpha,\beta)$ distributed.

\subsection{Student's t distribution}

If $x$ and $x_1$, ..., $x_n$ are independent Gaussian random variables with
mean $0$ and variance $1$, the quantity
\bq
t & = & \frac{x}{\sqrt{z/n}} \;\;\; \mbox{with} \;\;  
z = \sum\limits_{i=1}^n x_i^2,
\eq
is distributed according to a Student's t distribution with $n$
degrees of freedom.
The probability density function of Student's t distribution is given
by 
\bq
p(t) & = & \frac{1}{\sqrt{n \pi}} \frac{\Gamma \left( \frac{n+1}{2} \right)}
{ \Gamma \left( \frac{n}{2} \right)}
\left( 1 + \frac{t^2}{n} \right)^{- \frac{n+1}{2}}.
\eq
The range of $t$ is $-\infty < t < \infty$, and $n$ is not required to be an
integer.
For $n \ge 3$ Student's t distribution has mean $0$ and variance $n/(n-2)$.
To generate a random variable $t$ distributed according to 
Student's t distribution for $n > 0$
one generates a random variable $x$ distributed as Gaussian with mean $0$
and variance $1$, as well as a random variable $y$, distributed as according
to a gamma distribution with $k=n/2$ and $\lambda=1$. Then
\bq
t & = & x \sqrt{\frac{2n}{y}} 
\eq
is distributed as a $t$ with $n$ degrees of freedom.\\
\\
In the case $n=1$ Student's t distribution reduces to the Breit-Wigner
distribution:
\bq
p(t) & = & \frac{1}{\pi} \frac{1}{1+t^2}.
\eq
In this case it it simpler to generate the distribution as follows: Generate
to uniform random numbers $u_1$, $u_2$, set $v_1=2u_1-1$ and $v_2=2u_2-1$.
If $v_1^2+v_2^2\le 1$ set $t=v_1/v_2$, otherwise start again.

\end{appendix}

\end{document}